\documentclass[notitlepage,aps,floats,showpacs,showkeys,preprintnumbers,nofootinbib,11pt]{revtex4-1}
\pdfoutput=1

\usepackage{amsmath}
\usepackage{amssymb}
\usepackage{epsfig}
\usepackage{bbm}

\usepackage[pdftex,colorlinks=false,urlcolor=blue,linkcolor=blue]{hyperref}

\def\wtilde{\widetilde}
\def\wtil{\widetilde}
\def\mueff{\mu_{\text{eff}}}
\def\fbi{~{\rm fb}^{-1}}

\def\fb{~{\rm fb}}
\def\pb{~{\rm pb}}

\def\ie{{\it i.e.}}

\def\anti{\overline}
\def\sigsi{\sigma_{\rm SI}}

\def\mw{m_W}
\def\gev{~{\rm GeV}}
\def\tev{~{\rm TeV}}
\def\beq{\begin{equation}}
\def\eeq{\end{\equation}}
\def\bea{\begin{eqnarray}}
\def\eea{\end{eqnarray}}
\def\gam{\gamma}
\def\kap{\kappa}
\def\lam{\lambda}
\def\akap{A_\kap}
\def\alam{A_\lam}
\def\mhalf{m_{1/2}}

\def\mhusq{m_{H_u}^2}
\def\mhdsq{m_{H_d}^2}
\def\mssq{m_{S}^2}
\def\omghsq{\Omega h^2}
\def\amu{a_\mu}
\def\damu{\delta \amu}
\def\br{{\rm BR}}

\def\hsm{h_{\rm SM}}

\def\wpm{W^\pm}
\def\wp{W^+}
\def\wm{W^-}
\def\to{\rightarrow}
\def\hi{h_1}
\def\hii{h_2}
\def\hiii{h_3}
\def\hpm{H^\pm}
\def\hp{H^+}
\def\hm{H^-}
\def\ai{a_1}
\def\aii{a_2}
\def\mhi{m_{\hi}}
\def\mhii{m_{\hii}}
\def\mai{m_{\ai}}
\def\maii{m_{\aii}}
\def\mhiii{m_{\hiii}}
\def\mhpm{m_{\hpm}}

\def\cnone{\widetilde \chi_1^0}
\def\cpone{\widetilde \chi_1^+}
\def\cpmone{\widetilde \chi_1^\pm}
\def\mcnone{m_{\cnone}}

\def\mcpmone{m_{\cpmone}}

\def\cpmtwo{\widetilde \chi_2^\pm}

\def\mcpmtwo{m_{\cpmtwo}}
\def\cnthree{\widetilde \chi_3^0}
\def\mcnthree{m_{\cnthree}}
\def\cnfour{\widetilde \chi_4^0}
\def\mcnfour{m_{\cnfour}}
\def\cnfive{\widetilde \chi_5^0}
\def\mcnfive{m_{\cnfive}}
\def\hsm{h_{\rm SM}}
\def\mhsm{m_{\hsm}}
\def\msq{m_{\tilde q}}
\def\mgl{m_{\tilde g}}
\def\stopi{\tilde t_1}
\def\stopii{\tilde t_2}
\def\mstopii{m_{\stopii}}
\def\mstopi{m_{\stopi}}

\def\staui{\wtilde \tau_1}

\def\mstaui{m_{\staui}}
\def\tanb{\tan\beta}
\def\cotb{\cot\beta}
\def\tauptaum{\tau^+\tau^-}
\def\beq{\begin{equation}}
\def\eeq{\end{equation}}

\def\sig{\sigma}

\def\wwww{R^h_{VBF}(WW)}
\def\ggww{R^h_{gg}(WW)}
\def\wwpp{R^h_{VBF}(\gam\gam)}
\def\ggpp{R^h_{gg}(\gam\gam)}
\def\wwbb{R^h_{VBF}(bb)}
\def\ggbb{R^h_{gg}(bb)}

\def\wwppi{R^{\hi}_{VBF}(\gam\gam)}
\def\ggppi{R^{\hi}_{gg}(\gam\gam)}
\def\wwbbi{R^{\hi}_{VBF}(bb)}
\def\ggbbi{R^{\hi}_{gg}(bb)}

\def\wwppii{R^{\hii}_{VBF}(\gam\gam)}
\def\ggppii{R^{\hii}_{gg}(\gam\gam)}
\def\wwbbii{R^{\hii}_{VBF}(bb)}
\def\ggbbii{R^{\hii}_{gg}(bb)}
\def\vev#1{\langle #1 \rangle}
\def\rts{\sqrt s}
\def\lsim{\mathrel{\raise.3ex\hbox{$<$\kern-.75em\lower1ex\hbox{$\sim$}}}}
\def\gsim{\mathrel{\raise.3ex\hbox{$>$\kern-.75em\lower1ex\hbox{$\sim$}}}}
\def\epem{e^+e^-}
\def\mupmum{\mu^+\mu^-}

\def\bit{\begin{itemize}}
\def\eit{\end{itemize}}
\def\bec{\begin{center}}
\def\eec{\end{center}}
\def\what{\widehat}

\begin{document}
\title{Higgs Bosons at 98 and 125 GeV at  LEP and the LHC}

\author{Genevi\`eve~B\'elanger}\email{belanger@lapp.in2p3.fr}
\affiliation{LAPTH, Universit\'e de Savoie, CNRS, B.P.110, F-74941 Annecy-le-Vieux Cedex, France}
\author{Ulrich~Ellwanger}\email{Ulrich.Ellwanger@th.u-psud.fr}
\affiliation{Laboratoire de Physique Th\'eorique, Universit\'e Paris-Sud, Centre d'Orsay, F-91405 Orsay-Cedex, France}
\author{John F.~Gunion}\email{jfgunion@ucdavis.edu}
\author{Yun~Jiang}\email{yunjiang@ucdavis.edu}
\affiliation{\,Department of Physics, University of California, Davis, CA 95616, USA}
\author{Sabine~Kraml}\email{sabine.kraml@lpsc.in2p3.fr}
\affiliation{\,Laboratoire de Physique Subatomique et de Cosmologie, UJF Grenoble 1, CNRS/IN2P3, INPG, 
53 Avenue des Martyrs, F-38026 Grenoble, France}
\author{John~H.~Schwarz}\email{jhs@theory.caltech.edu}
\affiliation{Department of Physics, California Institute of Technology, Pasadena, CA 91125, USA}

\begin{abstract}
We discuss NMSSM scenarios in which the lightest Higgs boson $\hi$ is
consistent with the small LEP excess at $\sim 98\gev$ in $\epem\to Zh$
with $h\to b\anti b$ and  the heavier Higgs boson $\hii$ has the
primary features of the LHC Higgs-like signals at $125\gev$, including
an enhanced $\gam\gam$ rate. Verification or falsification of the
$98\gev$ $\hi$ may be possible at the LHC during the $14\tev$ run. The
detection of the other NMSSM Higgs bosons at the LHC and future
colliders is also discussed, as well as dark matter properties of the
scenario under consideration.

\end{abstract}

\preprint{UCD 2012-1, LPT Orsay 12-100, LAPTH-045/12, CALT-68-2888, LPSC12285}

\maketitle

\clearpage
\section{Introduction}

Data from the ATLAS and CMS collaborations~\cite{atlashiggs,cmshiggs}
provide an essentially $5\sigma$ signal for a Higgs-like resonance, $h$, with mass of order $125\gev$. 
Meanwhile, the CDF and D0 experiments have announced new results \cite{newtevatron}, based mainly on $Vh$ associated production with $h\to b\anti b$, that support the $\sim 125\gev$ Higgs-like signal.
While it is certainly possible that the observed signals in the various production/decay channels will converge towards their respective Standard Model (SM) values, the current central values for the signal strengths in individual channels deviate by about 1--2$\,\sigma$ from predictions for the $\hsm$. 
One of the most significant deviations in the current data is the enhancement in the $\gam\gam$ final state for both gluon fusion ($gg$) and vector boson fusion (VBF) production. Such a result is not atypical of models with multiple Higgs bosons in which the $b\anti b$ partial width  of the observed $h$ is reduced through mixing with a second (not yet observed at the LHC) Higgs boson, $h'$, thereby enhancing the $\gam\gam$ branching ratio of the $h$~\cite{Carena:2002qg,Ellwanger:2011aa,Cao:2012fz,Ellwanger:2012ke,Gunion:2012gc,Cao:2012yn}.
  In such models, a particularly interesting question is whether one could simultaneously explain the LHC signal and the small ($\sim 2\sigma$) LEP excess in $\epem\to Zb\anti b$ in the vicinity of $M_{b\anti b}\sim 98\gev$ \cite{Schael:2006cr,Barate:2003sz} using the $h'$ with $m_{h'}\sim 98\gev$. We recall that the LEP excess is clearly inconsistent with a SM-like Higgs boson at this mass, being only about $10-20\%$ of the rate predicted for the $\hsm$.  Consistency with such a result for the $h'$ is natural if the $h'$ couples at a reduced level to $ZZ$, which, in turn, is automatic if the $h$ has substantial $ZZ$ coupling, as required by the observed LHC signals.

In this paper we demonstrate that the two lightest CP-even Higgs bosons~\footnote{We assume absence of CP-violating phases in the Higgs sector.}, $\hi$ and $\hii$, of the Next-to-Minimal Supersymmetric Model (NMSSM) could have properties such that the $\hi$  fits the LEP excess at $\sim 98\gev$ while the $\hii$ is reasonably consistent with the Higgs-like LHC signals at $\sim 125\gev$, including in particular the larger-than-SM signal in the $\gam\gam$ channel.  
The NMSSM~\cite{Ellwanger:2009dp} is very attractive since it solves the $\mu$ problem of the minimal supersymmetric extension of the SM (MSSM): the ad hoc parameter $\mu$ appearing in the MSSM superpotential term $\mu \hat H_u \hat H_d$ is  generated in the NMSSM from the $\lam \hat S \hat H_u \hat H_d$ superpotential term when the scalar component $S$ of $\hat S$ develops a VEV $\vev{S}=s$: $\mueff=\lam s$. The three CP-even Higgs fields, contained in $H_u$, $H_d$ and $S$, mix and yield the mass eigenstates $\hi$, $\hii$ and $\hiii$.  
A $125\gev$ Higgs state with enhanced  $\gam\gam$ signal rate is easily obtained for large $\lam$ and small $\tanb$~\cite{Ellwanger:2011aa} (see also \cite{Ellwanger:2012ke,Gunion:2012gc}).  
To describe the LEP and LHC data the $\hi$ and $\hii$ must have $\mhi\sim 98\gev$ and $\mhii\sim 125\gev$, respectively, with the $\hi$ being largely singlet and the $\hii$ being primarily doublet (mainly  $H_u$ for the scenarios we consider).  In addition to the CP-even states, there are also two CP-odd states, $\ai$ and $\aii$, and a charged Higgs boson, $\hpm$. Verification of the presence of the three CP-even Higgs bosons and/or two CP-odd Higgs bosons would establish a Higgs field structure that goes beyond the two-doublet structure of the MSSM.

\section{Higgs Boson Production and Decay}

The main production/decay channels relevant for current LHC data are gluon fusion ($gg$)  and vector boson fusion (VBF)  
with Higgs decay to $\gam\gam$ or $ZZ^*\to 4\ell$. The LHC also probes $W,Z+$Higgs with Higgs decay to $b\anti b$, a channel for which Tevatron data is relevant, and $WW\to$Higgs with Higgs$\to\tau^+\tau^-$. 
We compute the ratio of the 
$gg$ or VBF induced Higgs cross section times the Higgs branching ratio to a given final state $X$, relative to the corresponding value for the SM Higgs boson, as 
\beq
   R^{h_i}_{gg}(X)\equiv {\Gamma( h_i\to gg) \ \br(h_i\to X)\over \Gamma(\hsm\to gg)\ \br(\hsm\to X)}, \quad 
   R^{h_i}_{\rm VBF}(X)\equiv {\Gamma(h_i\to WW) \ \br(h_i\to X)\over \Gamma(\hsm\to WW)\ \br(\hsm\to X)},
\eeq 
where $h_i$ is the $i^{th}$ NMSSM scalar Higgs, and $\hsm$ is the SM Higgs boson, taking $\mhsm=m_{h_i}$.
In the context of any two-Higgs-doublet plus singlets model, not all the $R^{h_i}$ are independent. For example,
$R^{h_i}_{VH}(X)=R^{h_i}_{VBF}(X)$, $R^{h_i}_{Y}(\tau\tau)=R^{h_i}_{Y}(bb)$~\footnote{This equality is altered by radiative corrections at large $\tanb$; however, these are small in our scenarios all of which have small to moderate $\tanb$ values.}  and $R^{h_i}_{Y}(ZZ)=R^{h_i}_{Y}(WW)$. A complete independent set of $R^{h_i}$'s can be taken to be (with $h=\hi$ or $h=\hii$)
\beq
   \ggww,\quad \ggbb,\quad\ggpp,\quad \wwww,\quad \wwbb,\quad \wwpp\,.
\eeq

In order to display the ability of the NMSSM to simultaneously explain the LEP and LHC Higgs-like signals, we turn to NMSSM scenarios with semi-unified GUT scale soft-SUSY-breaking. By ``semi-unified'' we mean universal gaugino mass parameter $\mhalf$, scalar (sfermion) mass parameter $m_0$, and trilinear coupling $A_0\equiv A_t=A_b=A_\tau$ at the GUT scale, but $\mhusq$, $\mhdsq$ and $\mssq$ as well as $\alam$ and $\akap$ are taken as non-universal at $M_{\rm GUT}$. Specifically, we use points from scans performed using NMSSMTools\,3.2.0~\cite{Ellwanger:2004xm,Ellwanger:2005dv,nmweb}, which includes the scans of~\cite{Gunion:2012gc} supplemented by additional runs following the same procedure as well as specialized MCMC chain runs designed to focus on parameter regions of particular interest.
All  the accepted points correspond to scenarios that obey all experimental constraints (mass limits and flavor constraints as implemented in NMSSMTools, 
$\omghsq<0.136$ and 2011 XENON100 constraints on the spin-independent scattering cross section) 
except that the SUSY contribution to the anomalous magnetic moment of the muon, 
$\damu$, is too small to explain the discrepancy between the observed value of $a_\mu$~\cite{Gray:2010fp}
  and that predicted by the SM. For a full discussion of the kind of NMSSM model employed see~\cite{Gunion:2012gc,Ellwanger:2012ke,Gunion:2012zd}.  

\begin{figure}[t]\centering
\includegraphics[width=0.65\textwidth]{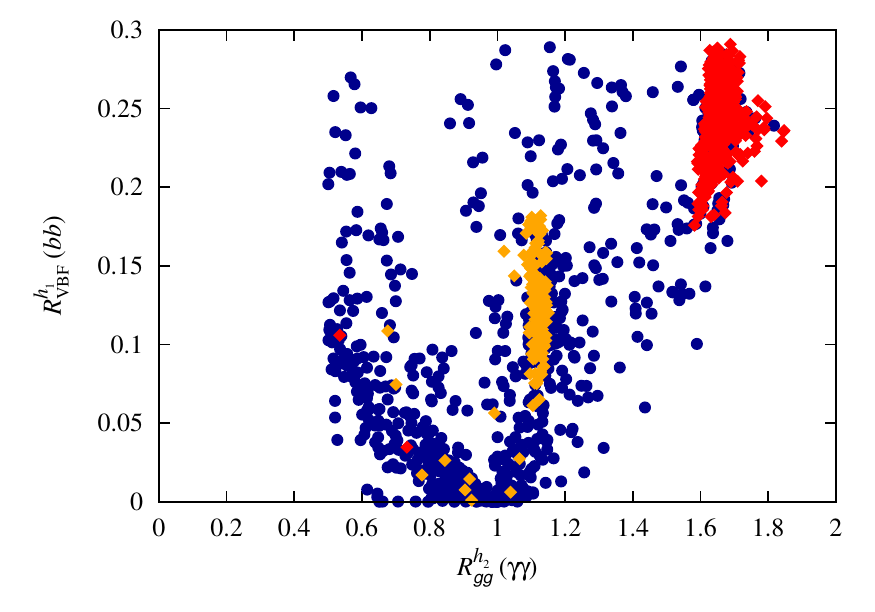}
\vspace*{-4mm}
\caption{Signal strengths (relative to SM) $\wwbbi$ versus $\ggppii$ for $\mhi\in[96,100]\gev$ and $ \mhii\in[123,128]\gev$.  In this and all subsequent plots, points with $\omghsq<0.094$ are represented by blue circles and points with $\omghsq\in[0.094,0.136]$ (the ``WMAP window") are represented by red/orange diamonds.  
\label{plot0}}
\end{figure}

We first display in Fig.~\ref{plot0} the crucial plot that shows $\wwbbi$ versus $\ggppii$  when $\mhi\in[96,100]\gev$ and $ \mhii\in[123,128]\gev$ are imposed in addition to the above mentioned experimental constraints.\footnote{Here the Higgs mass windows are designed to allow for theoretical errors in the computation of the 
Higgs masses.}  
(In this and all subsequent plots, points with $\omghsq<0.094$ are represented by blue circles and points with $\omghsq\in[0.094,0.136]$ (the ``WMAP window") are represented by red and orange diamonds. 
These two colors are associated with different LSP masses as will be discussed below.) 
Note that $\wwbbi$ values are required to be smaller than $0.3$ by virtue of the fact that the LEP constraint on the $\epem\to Z b\anti b$ channel with $M_{b\anti b}\sim 98\gev$ is included in the NMSSMTools program.  Those points with $\wwbbi$ between about $0.1$ and $0.25$ would provide the best fit to the LEP excess. (We note again that $\wwbbi$ is equivalent to $R^{\hi}_{V\hi}(b b)$ as relevant for LEP.) A large portion of such points have $\ggppii>1$ as preferred by LHC data. In all the remaining plots we will impose the additional requirements: $\ggppii>1$ and $0.1\leq\wwbbi\leq 0.25$. In the following,  we will refer to these NMSSM scenarios as the ``$98+125\gev$ Higgs scenarios".
To repeat, the $\ggppii>1$ requirement is such as to focus on points that could be consistent (within errors) with the enhanced $\gam\gam$ Higgs signal at the LHC of order 1.5 times the SM. The $0.1\leq \wwbbi\leq 0.25$ window is designed to reproduce the small excess seen in LEP data at $M_{b\anti b}\sim 98\gev$  in the $Z b\anti b$ final state.

\begin{figure}[t]\centering
\includegraphics[width=0.5\textwidth]{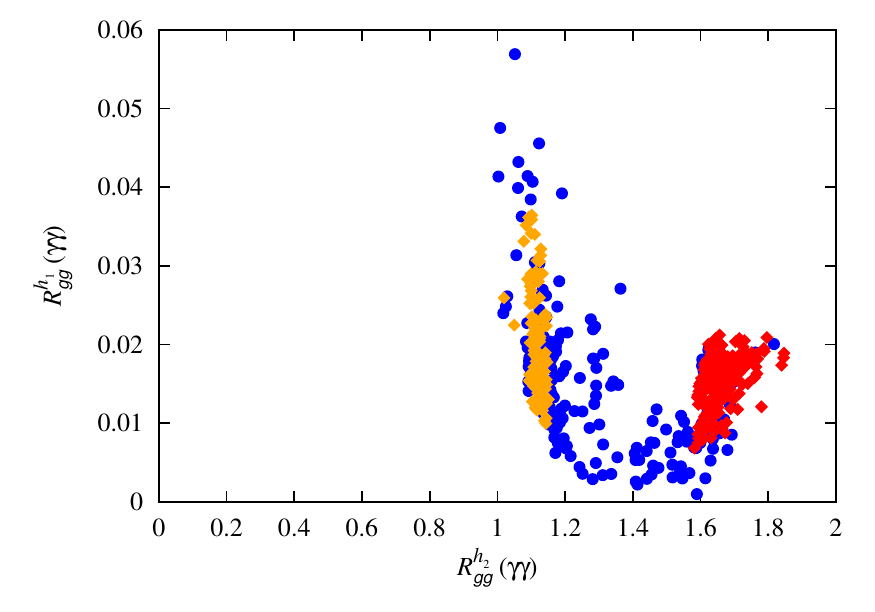}
\hspace{-5mm}
\includegraphics[width=0.5\textwidth]{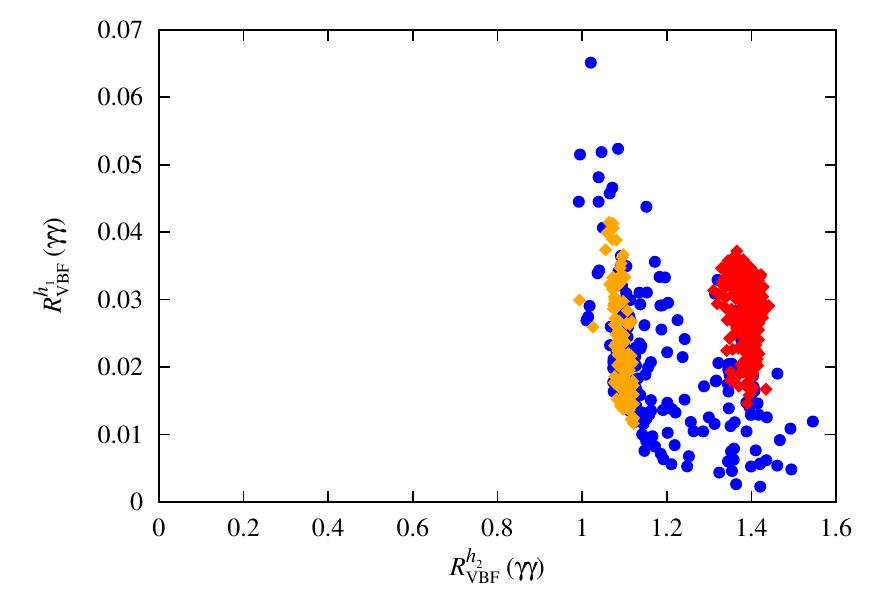}
\includegraphics[width=0.5\textwidth]{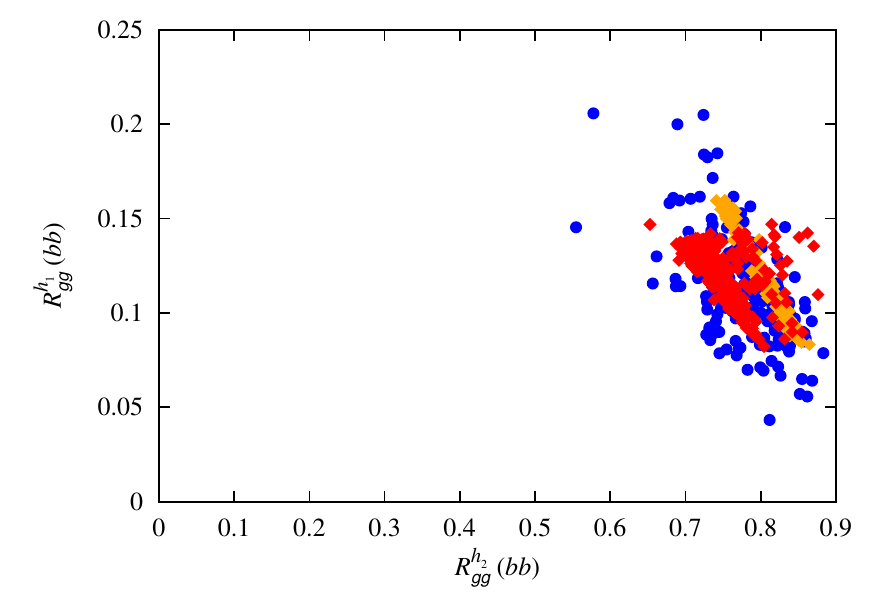}
\hspace{-5mm}
\includegraphics[width=0.5\textwidth]{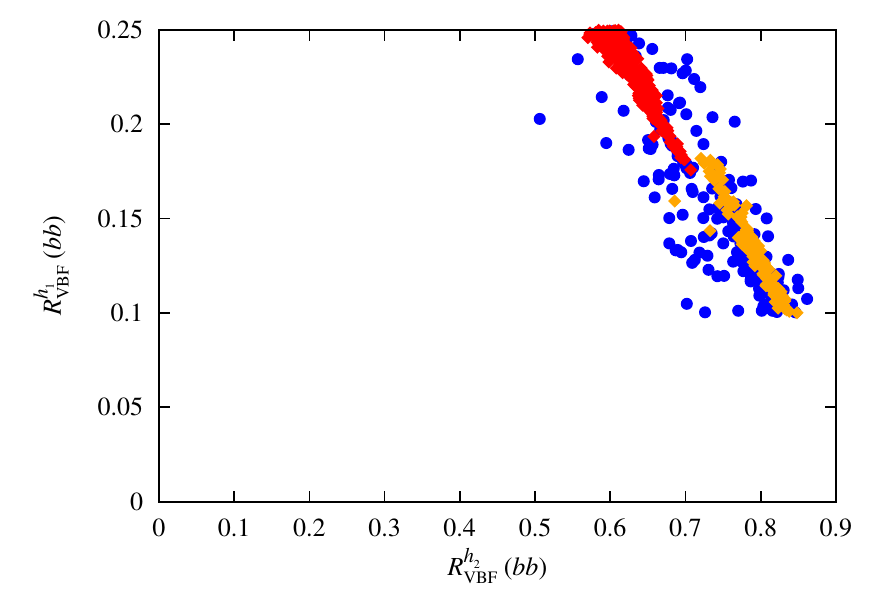}
\vspace*{-4mm}
\caption{For the $\hi$ and $\hii$, we plot (top) $\ggpp$ and $\wwpp$ and (bottom) $\ggbb$ and $\wwbb$ for NMSSM scenarios consistent with the LEP and LHC Higgs excesses. More specifically, in this and all subsequent plots we only show points that satisfy all the basic constraints specified in the text and that also satisfy $\mhi\in[96,100]\gev$, $\mhii\in[123,128]\gev$, $\ggppii>1$ and $\wwbbi\in[0.1,0.25]$. These we have termed the ``$98+125\gev$ Higgs scenarios".  
Regarding the WMAP-window points, we refer to the red diamonds as ``region A'' and to the orange ones as ``region B''. 
\label{plots1}}
\end{figure}

In Fig.~\ref{plots1}, we plot (upper row)  $\ggppi$ vs.\ $\ggppii$ and $\wwppi$ vs.\ $\wwppii$ and (lower row)  $\ggbbi$ vs.\ $\ggbbii$ and $\wwbbi$ vs.\ $\wwbbii$. In these and all subsequent plots, we only show points that satisfy all the basic constraints specified earlier and that also satisfy $\mhi\in[96,100]\gev$, $\mhii\in[123,128]\gev$, $\ggppii>1$ and $\wwbbi\in[0.1,0.25]$. The upper plots show that the $\hii$ can easily have an enhanced $\gam\gam$ signal for both $gg$ and VBF production whereas the $\gam\gam$ signal arising from the $\hi$ for both production mechanisms is quite small and unlikely to be observable. 
Note the two different $\ggppii$ regions for which $\omghsq$ lies in the WMAP window, one 
with $\ggppii\sim 1.6$ (region A, red diamonds)
and the other with $\ggppii\sim 1.1$ (region B, orange diamonds). As we will show later, region A 
corresponds to $\mcnone\sim 77\gev$ and $\mstopi$ between $197\gev$ and $1\tev$,  
while the region B corresponds to $\mcnone>93\gev$ and $\mstopi>1.8\tev$. 
These same two regions will emerge in many subsequent figures. If $\ggppii$ ends up converging to a large value, then masses for all strongly interacting SUSY particles would be close to current limits if the present $98+125\gev$ LEP-LHC Higgs scenario applies.

The bottom row of the figure focuses on the $b\anti b$ final state. We observe the reduced $\ggbbii$ and $\wwbbii$ values that are associated with reduced $b\anti b$ width (relative to the SM) needed to have enhanced $\ggppii$ and $\wwppii$.  Meanwhile, the $\ggbbi$ and $\wwbbi$ values are such that the $\hi$ could not yet have been seen at the Tevatron or LHC. Sensitivity to $\ggbbi$ ($\wwbbi$) values from 0.05 to 0.2 (0.1 to 0.25) will be needed at the LHC. This compares to expected sensitivities after the $\rts=8\tev$ run in these channels to $R$ values of at best $0.8$.\footnote{Here, we have used Fig.~12 of~\cite{cmshiggs} extrapolated to a Higgs mass near $98\gev$ and assumed $L=20\fbi$ each  for ATLAS and CMS.}  Statistically, a factor of 4 to 10 improvement requires integrated luminosity of order 16 to 100 times the current $L=10\fbi$.  Such large $L$ values will only be achieved after the LHC is upgraded to $14\tev$, although we should note that the luminosity required to probe this signal at 14 TeV could 
be lower than indicated by this simple estimate as the sensitivity to 
the Higgs signal improves at higher energies. Finally, the reader should note that for WMAP-window points the largest $\wwbbi$ values occur for region A described above for which supersymmetric particle masses are as small as possible.

\section{Other NMSSM particles and parameters}

\begin{figure}[b]\centering
\includegraphics[width=0.5\textwidth]{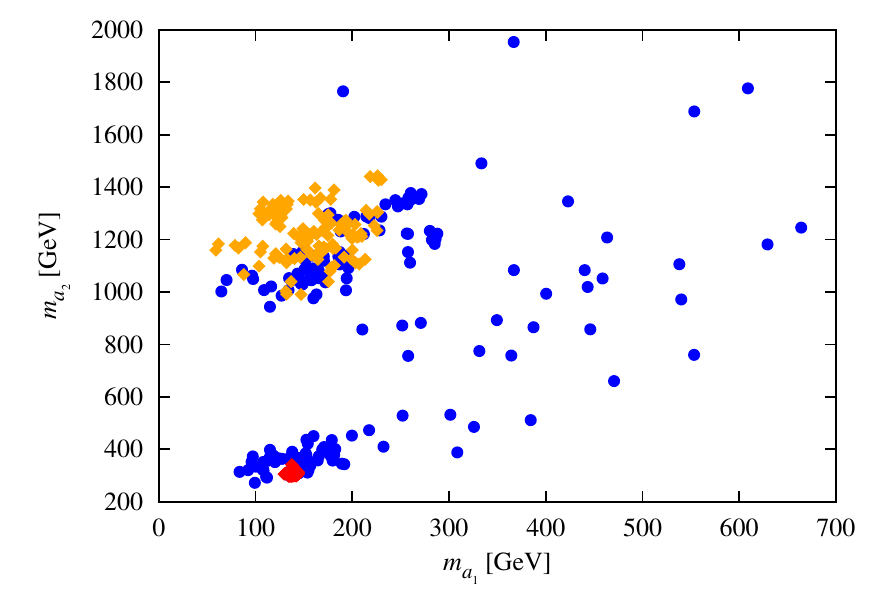}
\vspace*{-4mm}
\caption{Scatter plot of $\maii$ versus $\mai$ for the 98+125 GeV scenario; note that $\maii\simeq\mhiii\simeq\mhpm$. Note that in this figure there is a dense region, located at $(\mai,\maii)\sim (130,330)\gev$, of strongly overlapping red diamond points. These are the points associated with the low-$\mcnone$ WMAP-window region of parameter space.  Corresponding dense regions appear  in Figs. \ref{plots3} -- \ref{plots4} and \ref{cdeffvsm}. \label{plots2}}
\end{figure}

\begin{figure}[t]\centering
\includegraphics[width=0.5\textwidth]{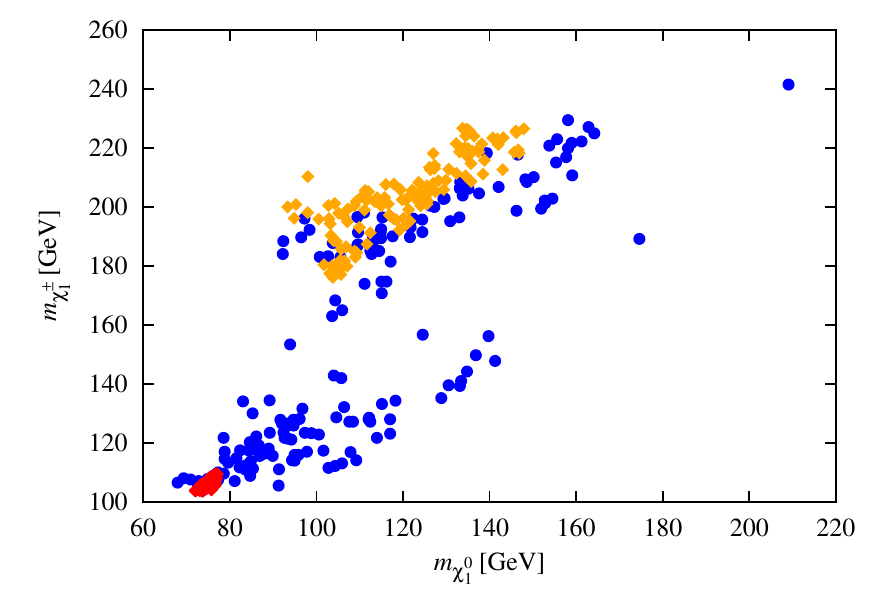}
\hspace{-5mm}
\includegraphics[width=0.5\textwidth]{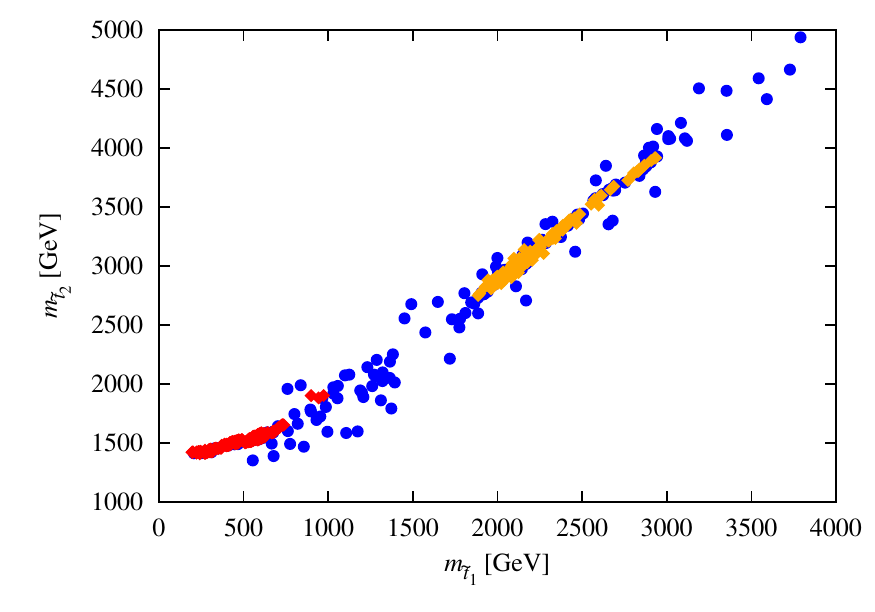}
\includegraphics[width=0.5\textwidth]{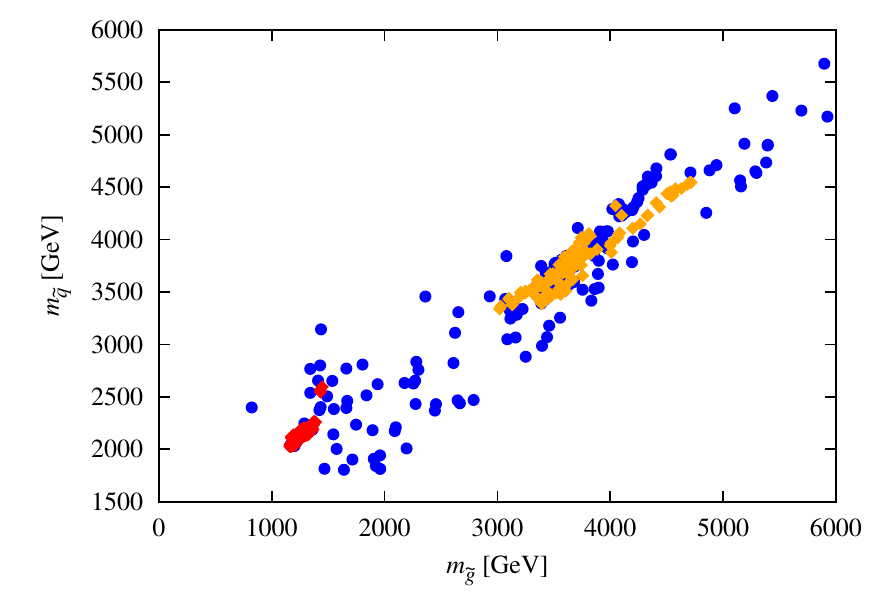}
\hspace{-5mm}
\includegraphics[width=0.5\textwidth]{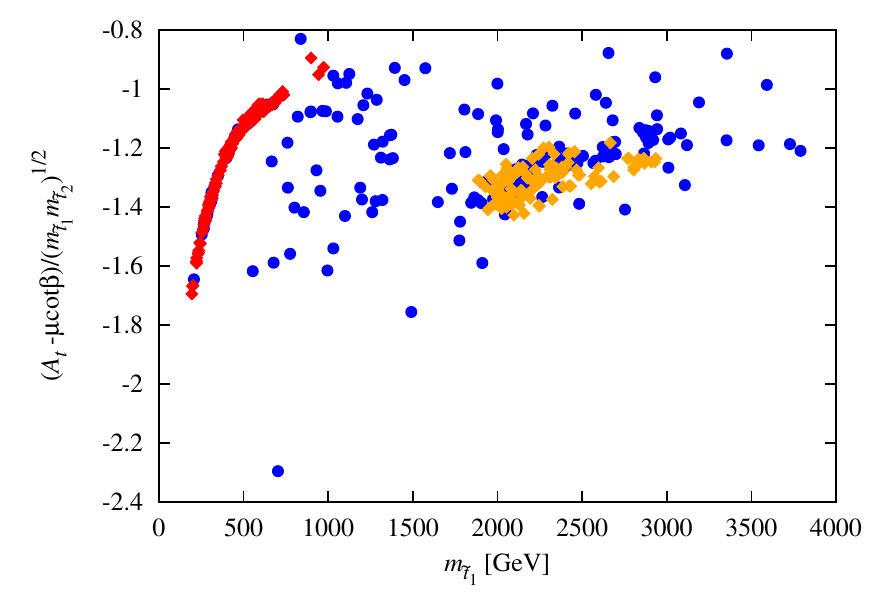}
\includegraphics[width=0.5\textwidth]{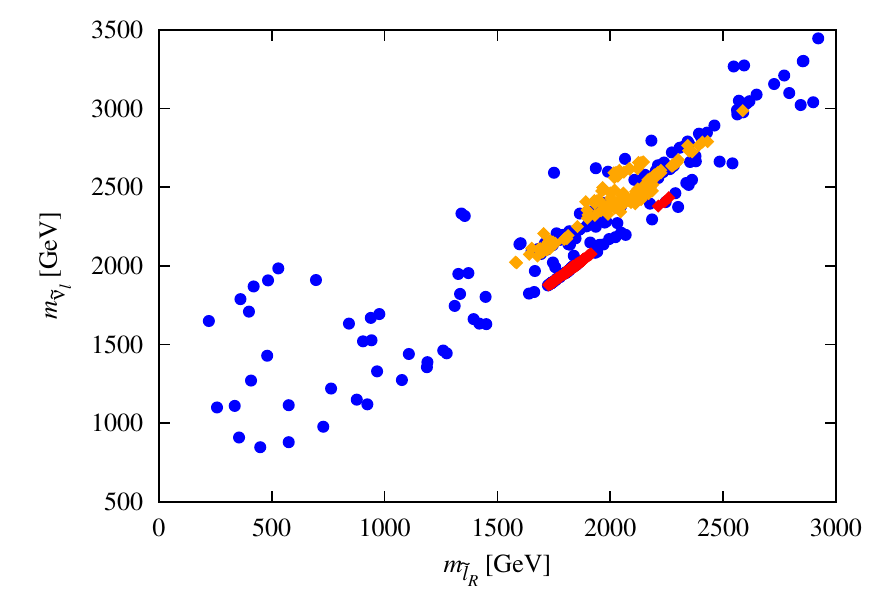}
\hspace{-5mm}
\includegraphics[width=0.5\textwidth]{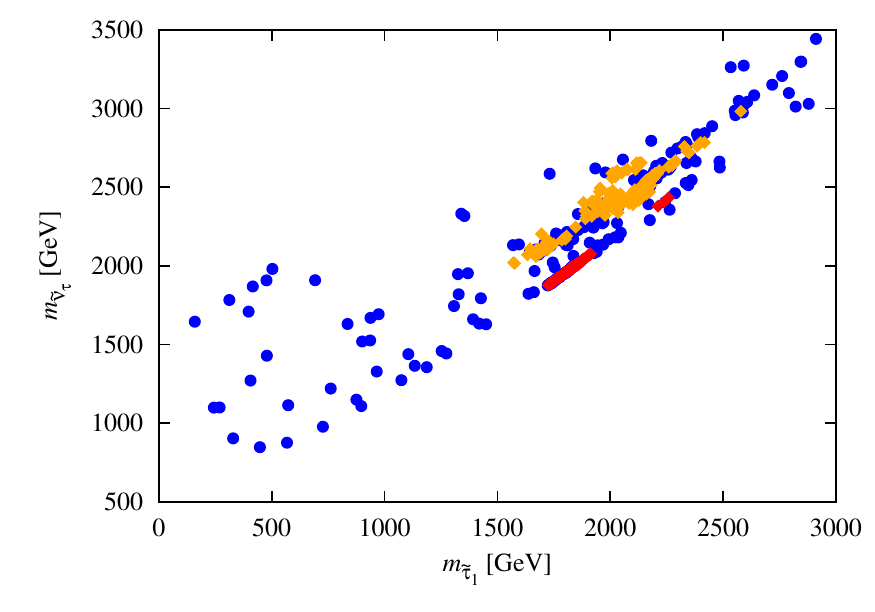}
\caption{Plots showing $\mcnone$, $\mcpmone$, $\mstopi$, $\mstopii$, $\msq$, $\mgl$, and the mixing parameter
$(A_t-\mu\cotb)/\sqrt{\mstopi\mstopii}$. Also shown are $m_{\wtil \ell_R}$, $m_{\wtil \nu_{\ell}}$, $\mstaui$ and $m_{\wtilde \nu_\tau}$, where $\ell=e,\mu$.  \label{plots3}}
\end{figure}

\begin{figure}[t]\centering
\includegraphics[width=0.5\textwidth]{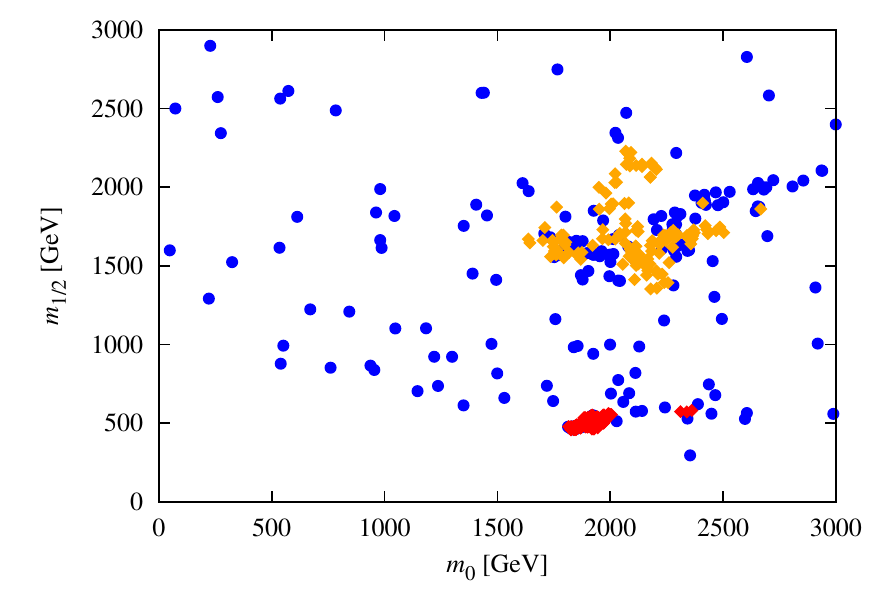}
\hspace{-5mm}
\includegraphics[width=0.5\textwidth]{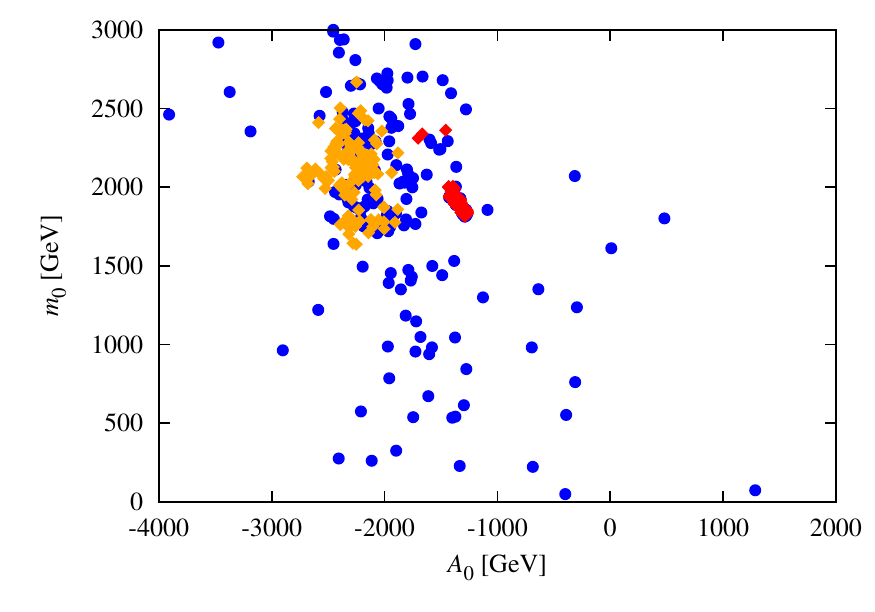}
\hspace{-5mm}
\includegraphics[width=0.5\textwidth]{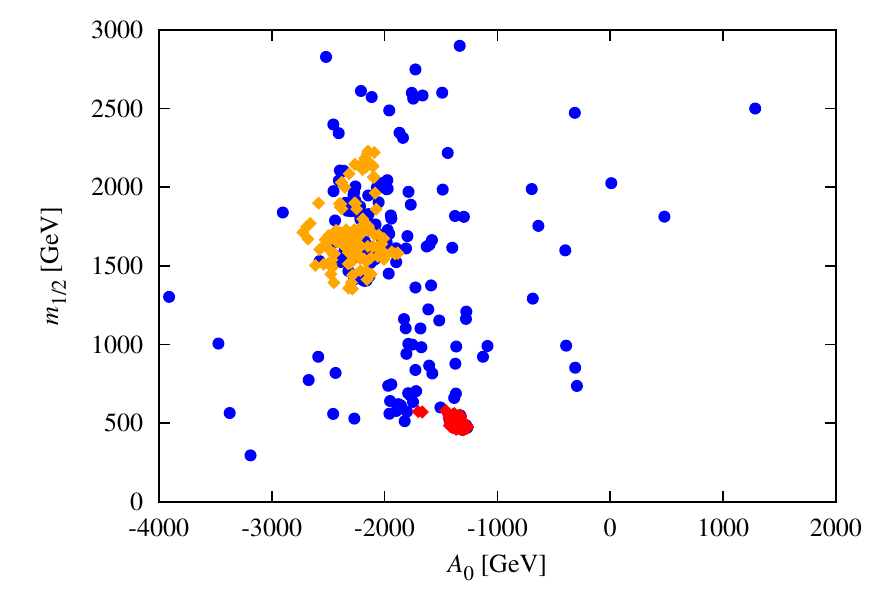}
\hspace{-5mm}
\includegraphics[width=0.5\textwidth]{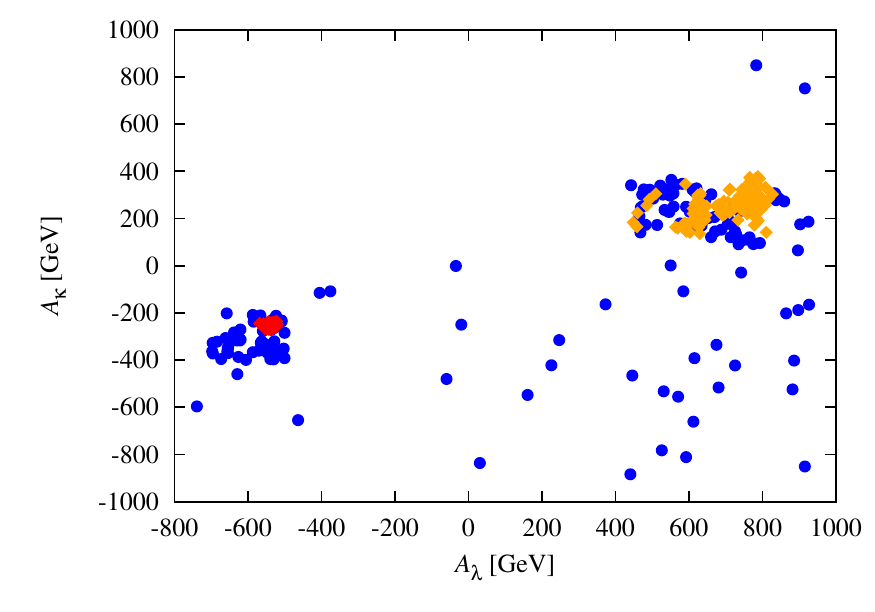}
\hspace{-5mm}
\includegraphics[width=0.5\textwidth]{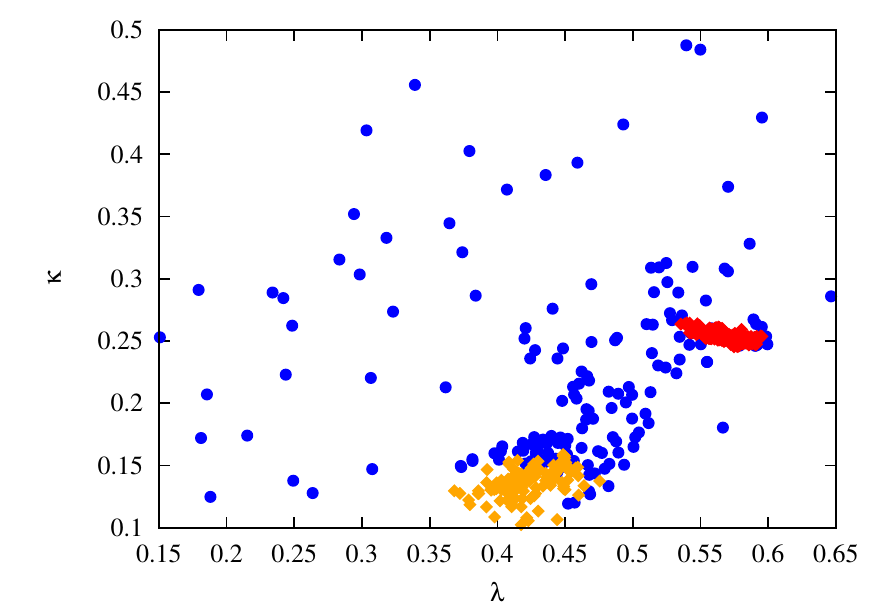}
\hspace{-5mm}
\includegraphics[width=0.5\textwidth]{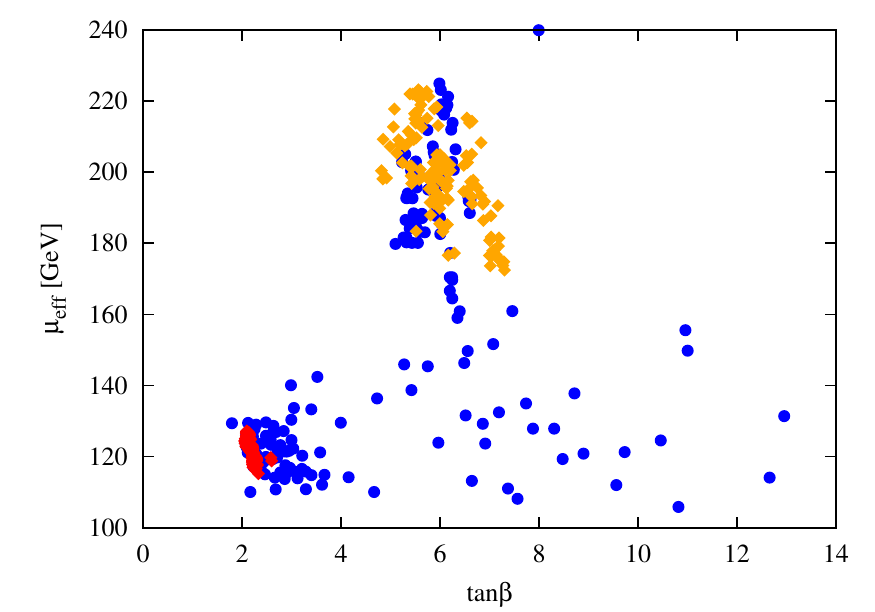}
\vspace*{-4mm}
\caption{GUT scale and SUSY scale parameters leading to the $98+125\gev$ LEP-LHC Higgs scenarios. \label{plots5}}
\end{figure}

It is also very interesting to consider expectations for the other NMSSM particles in these scenarios.  For this purpose, we present a series of plots. Figure~\ref{plots2} displays the pseudoscalar masses in the $\mai$--$\maii$ plane. We do not plot $\mhiii$ nor $\mhpm$ since their masses are such that $\mhiii\simeq \mhpm\simeq \maii$ for the scenarios considered. We note that small $\mai$ is typical of the WMAP-window points.    We discuss discovery prospects for the $\ai$ later in the paper.  
The masses of some crucial SUSY particles are displayed in Fig.~\ref{plots3}. We observe the typically low values of $\mcnone$ and $\mcpmone$, the possibility of $\mstopi$ as small as $197\gev$, the mostly modest values of the mixing parameter $(A_t-\mu\cotb)/\sqrt{\mstopi\mstopii}$, and the fact that the predicted $\msq$ and $\mgl$ are beyond current experimental limits, although the lowest values (as found in particular in region A) may soon be probed. Note that $\mgl$ can be below  $m_{\wtil \ell_R}$  (as common in constrained models when $m_0$ is large) for some points, including the points in region A.  Low values of $\mcnone$ are typical for the scan points, but more particular to this model are the rather low values of $\mcpmone$.  
ATLAS and CMS are currently performing analyses that could in principle be sensitive to the $\mcpmone$ values predicted in this model. For some points, $\mcpmone-\mcnone$ can be rather small, implying some difficulty in isolating the leptons or jets associated with $\cpmone\to \cnone+X$ decays. However, it should be noted that for the WMAP-window points $\mcpmone-\mcnone$ is typically quite substantial, at least $35\gev$ for the low-$\mcnone$ points,  so that for these points the above difficulty would not arise. Of particular interest is the very large range of $\mstopi$ that arises in the $98+125\gev$ LEP-LHC Higgs scenarios.  
For lighter values of $\mstopi$, as typical of the WMAP-window points in region A, the $\stopi$ always decays via $\stopi\to \cpone b$ or $\stopi\to \cnone t$, the latter being absent when $\mstopi<\mcnone+m_t$.  
At high $\mstopi$, these same channels are present but also $\stopi\to \wtil \chi^0_{2,3,4,5} t$ can  be important, which channels being present depending upon whether $\mstopi-m_{\chi^0_{2,3,4,5}}-m_t>0$ or not.

It is interesting to survey the GUT scale parameters that lead to the scenarios of interest.  Relevant plots are shown in Fig.~\ref{plots5}. No particular regions of these parameters appear to be singled out aside from some preference for negative values of $A_0$.  
These plots show clearly that  scenarios A and B correspond to distinct regions in the parameter space. Note however that the density  of red points in these plots is purely due to our scan procedures which have some focus on region A.

\section{Dark matter, including LSP and light chargino compositions}\label{sect:DM}

\begin{figure}[b]\centering
\includegraphics[width=0.5\textwidth]{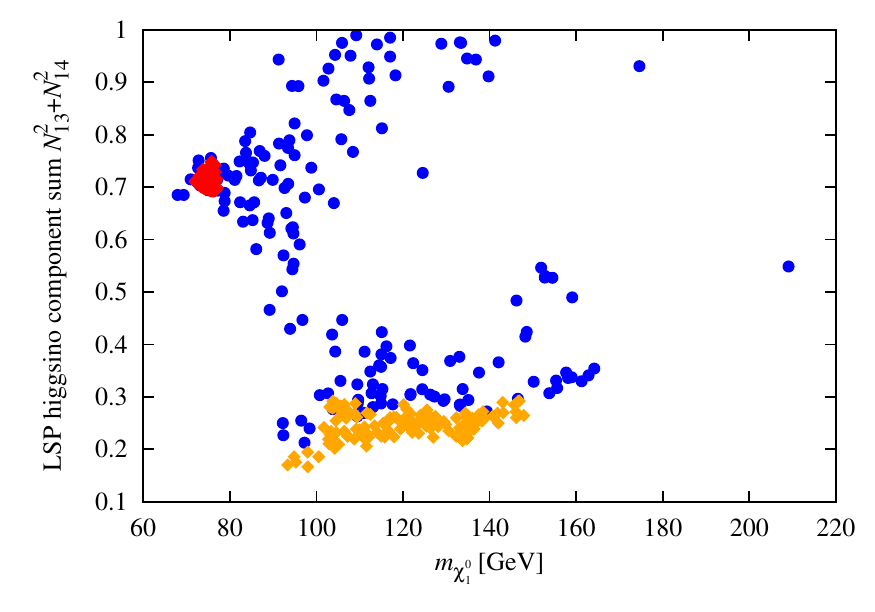}
\hspace{-5mm}
\includegraphics[width=0.5\textwidth]{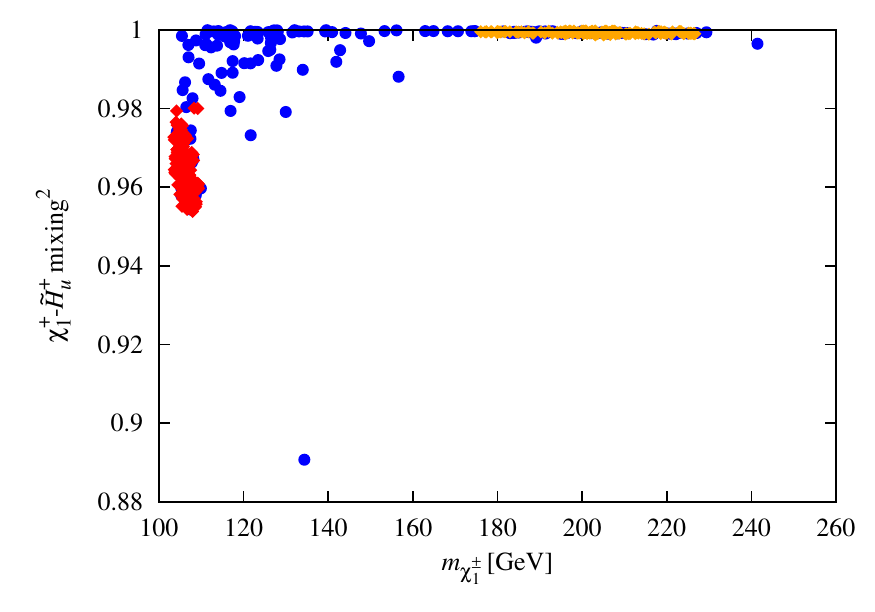}
\caption{Neutralino and chargino compositions for the $98+125\gev$ LEP-LHC Higgs scenarios.  \label{plots5.5}}
\end{figure}

\begin{figure}[b]\centering
\includegraphics[width=0.5\textwidth]{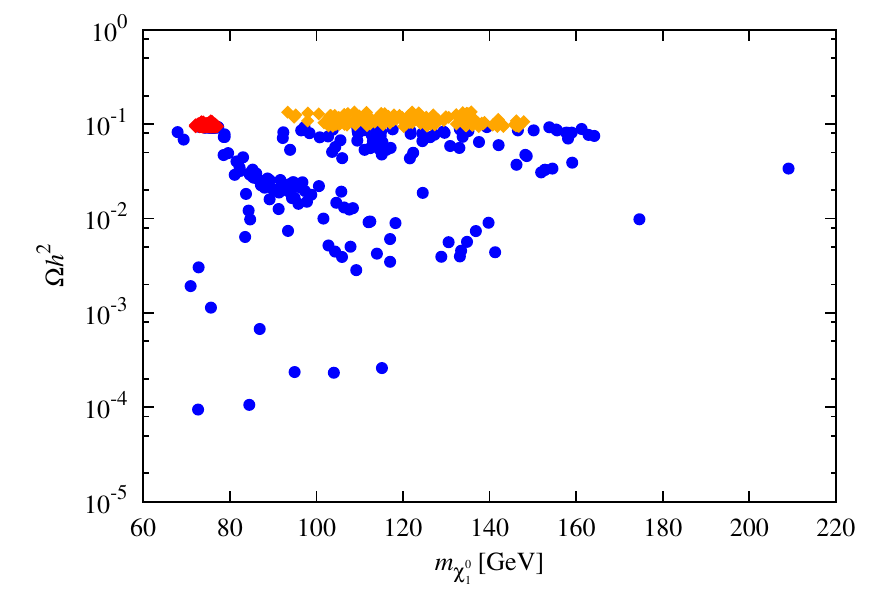}
\hspace{-5mm}
\includegraphics[width=0.5\textwidth]{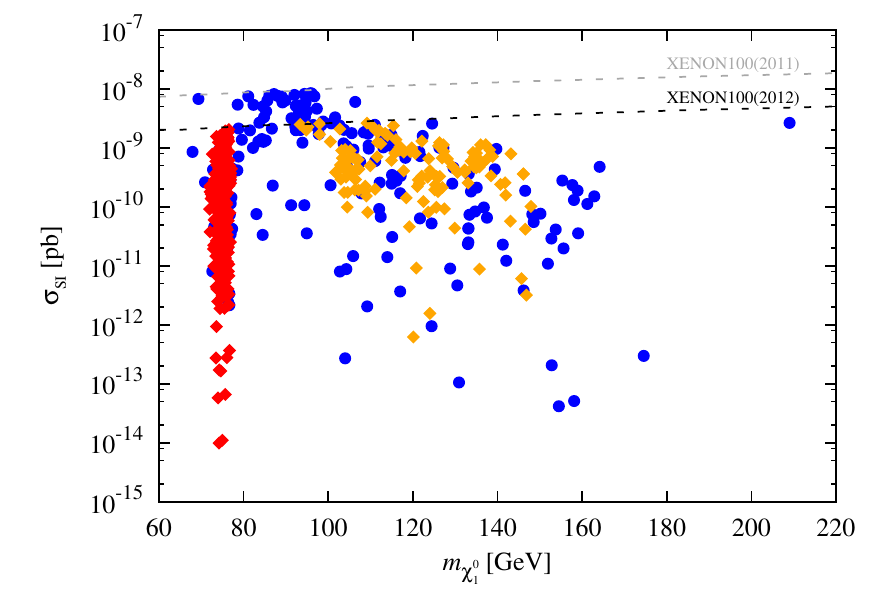}
\includegraphics[width=0.5\textwidth]{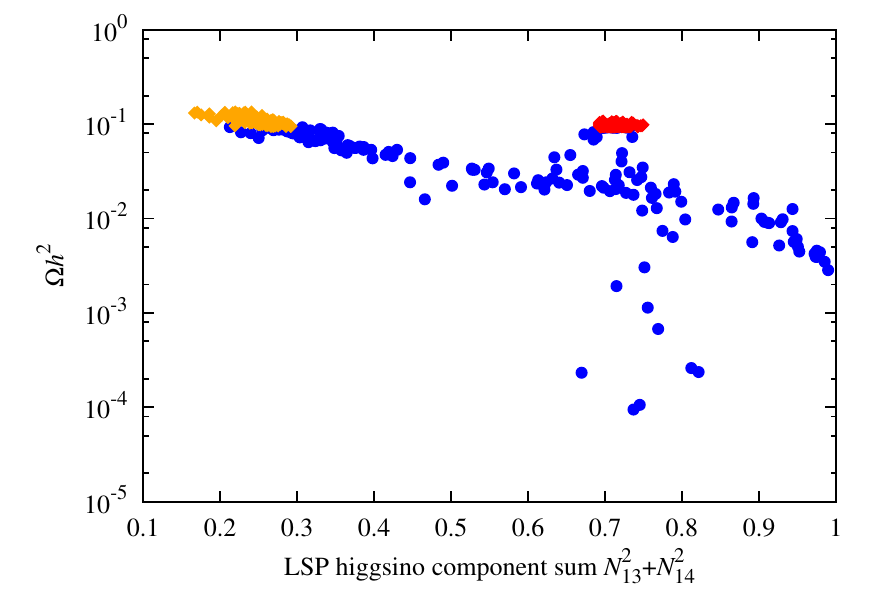}
\hspace{-5mm}
\includegraphics[width=0.5\textwidth]{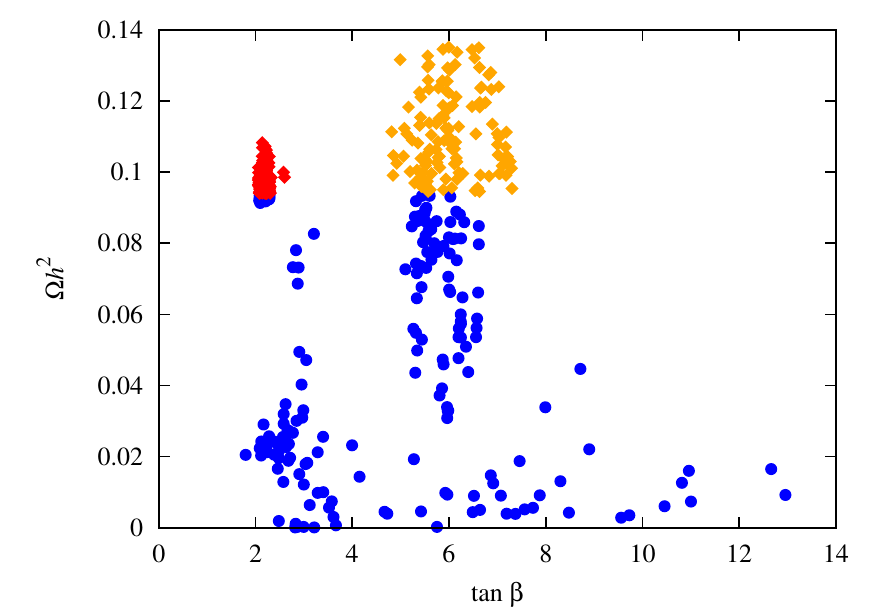}
\caption{Dark matter properties for the $98+125\gev$ LEP-LHC Higgs scenarios.    \label{plots4}}
\end{figure}

The composition of the $\cnone$ and the $\cpmone$ are crucial when it comes to the relic density of the $\cnone$.
For those points in the WMAP window in region A (red diamonds), the $\cnone$ can have a large Higgsino fraction since the $\cnone\cnone\to \wp\wm$ annihilation mode (mainly via $t$-channel exchange of the light Higgsino-like --- see second plot of Fig.~\ref{plots5.5} --- chargino) is below threshold; the group of points with $\mcnone>93\gev$ (region B, orange diamonds) can lie in the WMAP window only if the $\cnone$ does not have a large Higgsino fraction.
This division is clearly seen in Fig.~\ref{plots5.5}. We note that to a reasonable approximation the singlino fraction of the $\cnone$ is given by 1 minus the Higgsino fraction plotted in the left-hand window of the figure.

Dark matter (DM) properties for the surviving NMSSM parameter points are summarized in Fig.~\ref{plots4}.  
Referring to the figure, we see a mixture of blue circle points (those with  $\omghsq<0.094$) and red/orange diamond points (those with $0.094\leq \omghsq \leq 0.136$, \ie\ in the WMAP window).  The main mechanism at work to make $\omghsq$ too small for many points is rapid $\cnone\cnone$ annihilation to $W^+W^-$ due to a substantial Higgsino component of the $\cnone$ (see third plot of Fig.~\ref{plots4}).  Indeed, the relic density of a Higgsino LSP is typically of order $\omghsq\approx 10^{-3}-10^{-2}$.  
As the Higgsino component declines  $\omghsq$ increases 
and (except for the strongly overlapping points with $\mcnone<\mw$, for which $\cnone\cnone\to\wp\wm$ is below threshold) it is the points for which the LSP is dominantly singlino that have large enough  $\omghsq$ to fall in the WMAP window.

Also plotted in Fig.~\ref{plots4} is the spin-independent direct detection cross section, $\sigsi$, as a function of $\mcnone$.  First of all, we note that the 2012 XENON100 limits on $\sigsi$ are obeyed by all the points that have $\omghsq$ in the WMAP window, even though our scans only implemented the 2011 XENON100 limits --- indeed only a modest number of the $\omghsq<0.094$ points are inconsistent with the 2012 limits.  The $\sigsi$ plot also shows that experiments probing the spin-independent cross section will reach sensitivities that will probe some of the  $\sigsi$ values that survive the 2012 XENON100 limits relatively soon, especially the $\mcnone>93\gev$ points that are in the WMAP window (region B). However, it is also noteworthy that the $\mcnone\sim 75\gev$ points in region A can have very small $\sigsi$. 

The fourth plot of Fig.~\ref{plots4} and fifth  plot of Fig.~\ref{plots5} illustrate clearly the two categories of WMAP-window points. The first category (A)  of points is that for which the $\cnone$ has low mass and large Higgsino component with $\tanb\in [2,2.6]$ and $\lam\in[0.53, 0.6]$; ; the second category (B) is that for which
$\mcnone>93\gev$, $\tanb\in[5,7]$ and $\lam\in[0.37,0.48]$ .

\begin{figure}[h]\centering
\includegraphics[width=0.5\textwidth]{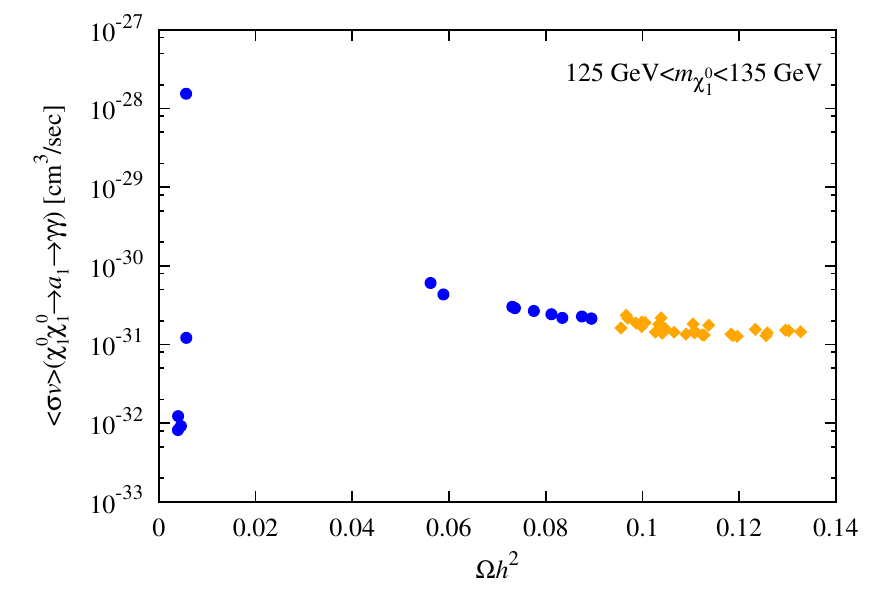}
\caption{We plot $\langle \sig v\rangle (\cnone\cnone\to \ai\to \gam\gam)$ vs. $\omghsq$ for just those points with $\mcnone\in [125,135]\gev$.  \label{plots4b}}
\end{figure}

It is interesting to discuss whether or not any of the $98+125\gev$ Higgs scenario  points are such as to describe the monochromatic signal at $130\gev$ observed in the Fermi-LAT data~\cite{Weniger:2012tx}.  We recall that the observation requires  $\langle \sig v\rangle (\cnone\cnone\to \gam\gam)\sim 10^{-27} {\rm cm}^3/{\rm sec}$ (this quoted value assumes standard dark matter density, $\rho\sim 0.3$).\footnote{Here, and below, $v$ is the very small velocity typical of dark matter in the current epoch,  $v\sim 10^{-3} c$, as relevant for indirect detection of the $\cnone$ through $\cnone\cnone$  annihilations.  This, of course, differs from the velocity at the time of freeze out, which is substantially higher.}
The situation is illustrated in Fig.~\ref{plots4b} where we plot  $\langle \sig v\rangle (\cnone\cnone\to \ai\to \gam\gam)$ vs. $\omghsq$ for just those points with $\mcnone\in [125,135]\gev$. (It is the $s$-channel $\ai$ diagram that can give a large $\vev{\sig v}$.) We observe that points with $\omghsq$ in the WMAP window have values of $\vev {\sigma v}$ four orders of magnitude below that required to explain the excess.  Those points with the largest $\vev{\sig v}$ always have quite small $\omghsq$ and hence $\rho_{DM}$. 
Incidentally, we have checked that all the points in our plots are fully consistent with the current bounds from the continuum $\gam$ spectrum as measured by Fermi-LAT ~\cite{Atwood:2009ez,Bringmann:2012vr}.

If the $130\gev$ gamma ray line is confirmed, then the above  questions will need to be explored more carefully. That a fully general NMSSM model (no GUT scale unifications) can be consistent simultaneously with the WMAP window, $\langle \sig v \rangle (\cnone\cnone\to\ai \to \gam\gam)\sim 10^{-27}{\rm cm}^3/{\rm sec}$, a Higgs mass close to $125\gev$ and 2011 XENON100 constraints was demonstrated in~\cite{Das:2012ys}. However, the value of $\mai$ has to be carefully tuned and the $125\gev$ Higgs couplings to all particles (including photons)  must be within 5\% of those for a SM Higgs boson of this mass, implying difficulty in describing the enhanced $\gam\gam$ LHC rates in this channel. Some general (non-NMSSM) theoretical discussions of the $130\gev$ line in the context of DM appear in \cite{Bai:2012qy,Bringmann:2012ez}.

\section{\boldmath Future tests of the $98+125\gev$  Higgs Scenario}

A critical issue is what other observations 
would either confirm or rule out the $98+125\gev$ LEP-LHC Higgs scenarios.  We first discuss possibilities at the LHC and then turn to future colliders, including  a future $e^+e^-$ collider, a possible $\gam\gam$ collider and a future $\mupmum$ collider.

\vspace*{-.2in}
\subsection{Direct Higgs production and decay at the LHC}
\vspace*{-.1in}

\begin{figure}[h!]
\vspace*{-.5in}
\hspace*{-7mm}
\includegraphics[width=1.05\textwidth]{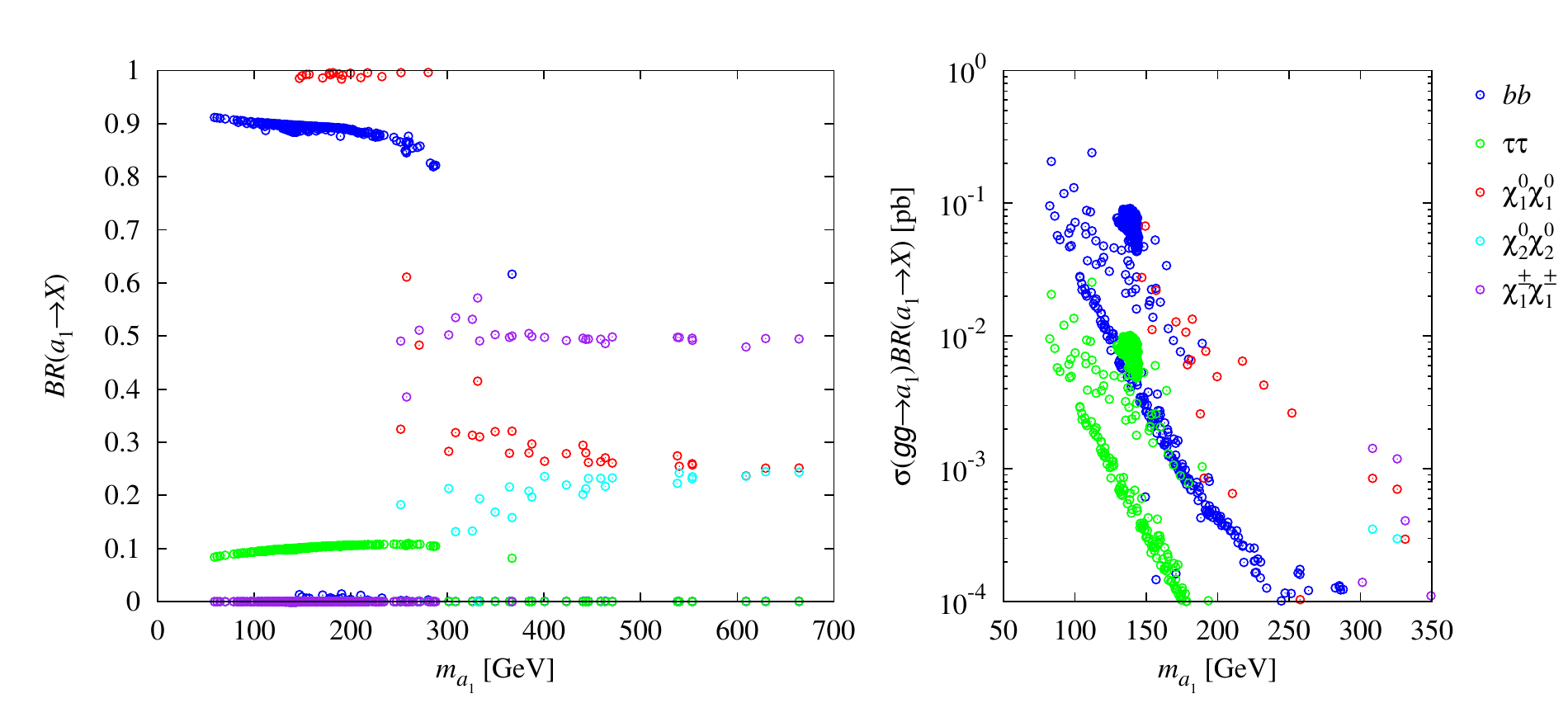}
\vskip -.25 in
\hspace*{-7mm}
\includegraphics[width=1.05\textwidth]{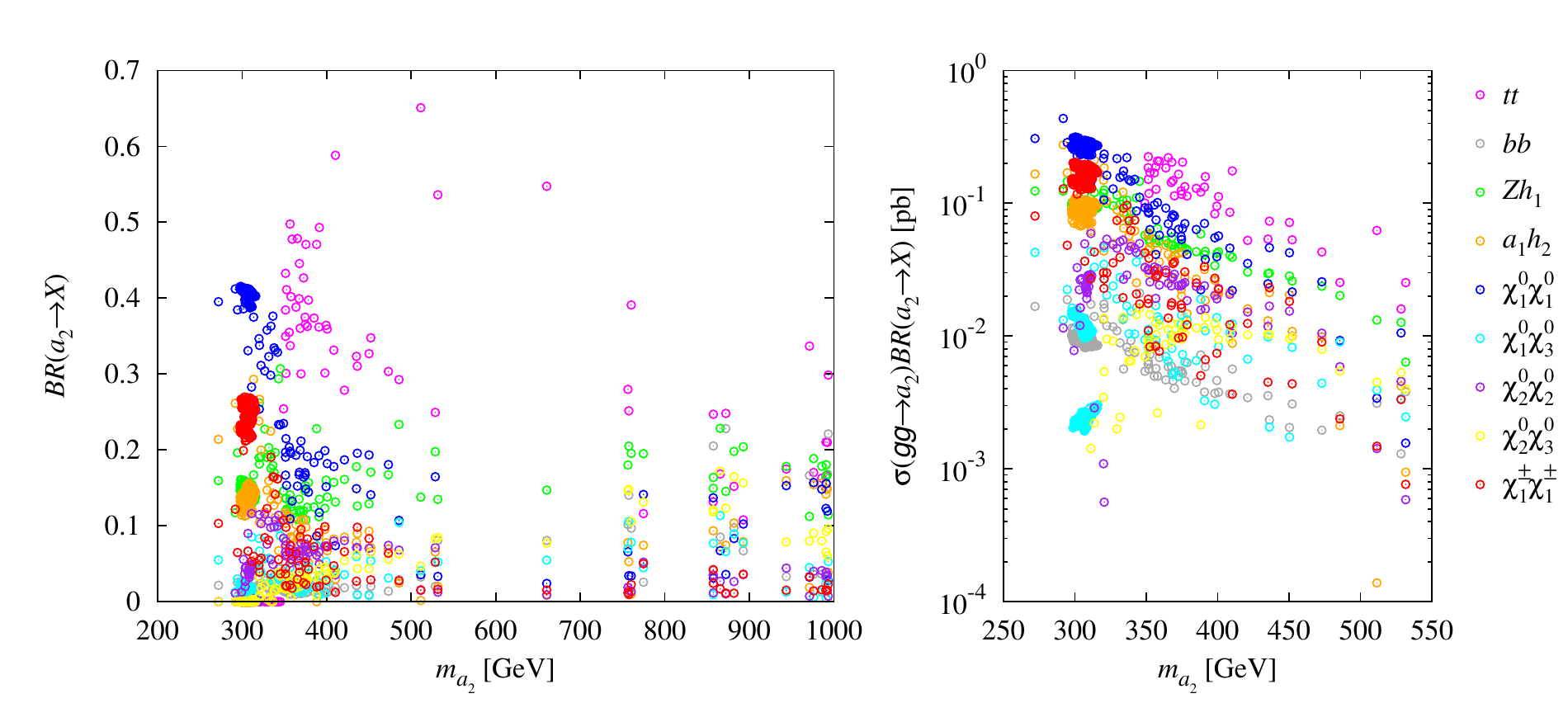}
\vskip -.25in
\hspace*{-7mm}
\includegraphics[width=1.05\textwidth]{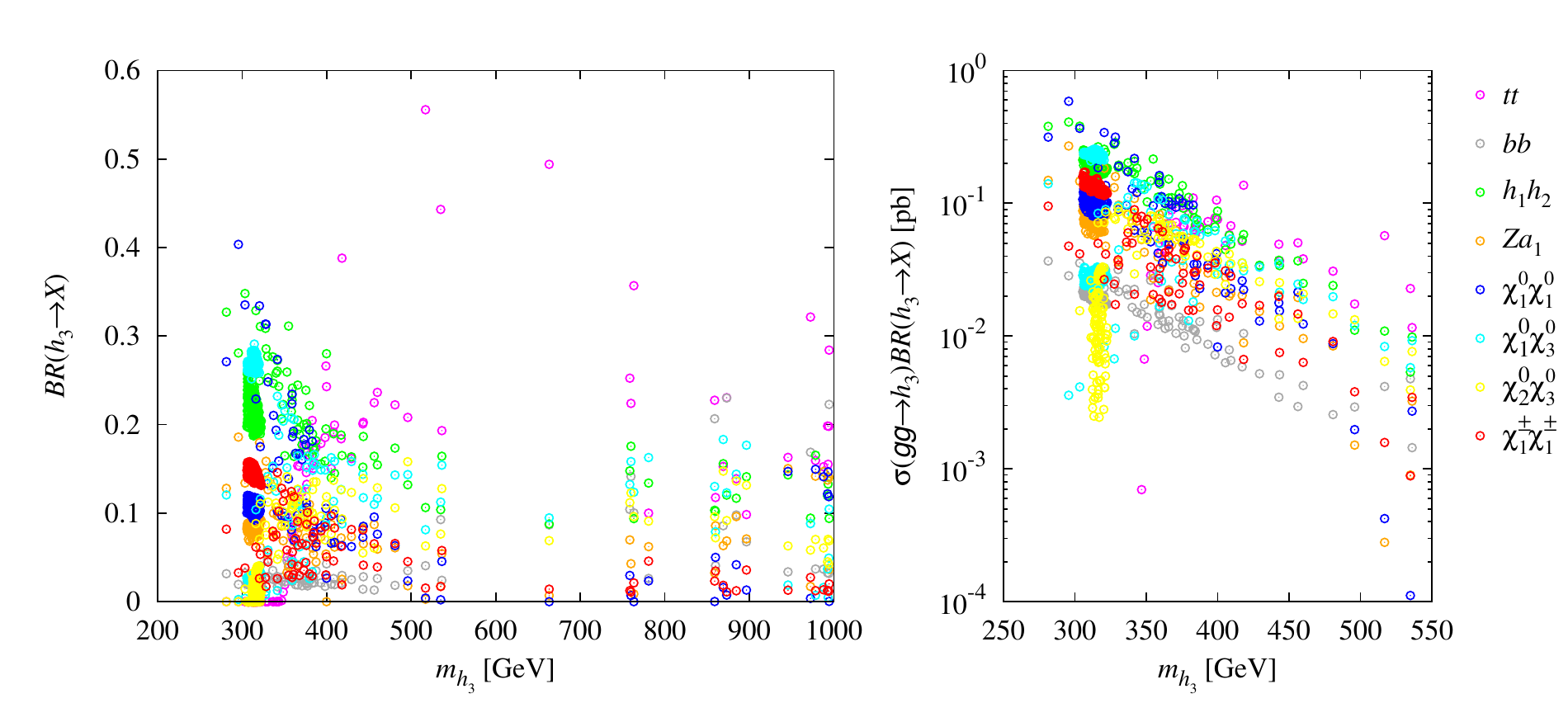}
\vspace*{-.5in}
\caption{Branching ratios and LHC cross sections in the $gg$ fusion mode (at $\rts=8\tev$) for $\ai$, $\aii$ and $\hiii$\label{otherhiggs}}
\end{figure}

We have already noted in the discussion of Fig.~\ref{plots1} that $gg$ and VBF production of the $\hi$ with $\hi\to b\anti b$ provide event rates that might eventually be observable at the LHC once much higher integrated luminosity is attained. Other possibilities include production and decay of the $\ai$, $\aii$, and $\hiii$. Decay branching ratios and LHC cross sections in the $gg$ fusion mode for $\ai$, $\aii$ and $\hiii$ are shown in Fig.~\ref{otherhiggs}. Since the $\ai$ is dominantly singlet in nature, its production rates at the LHC are rather small. The largest $\sig\br(X)$ values are in the $X=b\anti b$ final state, but this final state will have huge backgrounds. When allowed, $\sig\br(X)$ for $X=\cnone\cnone$  can be significant, but observation of this invisible final state would require a jet or photon tag that would further decrease the cross section.  The $\aii$ is dominantly doublet and provides better discovery prospects. If $\maii>2m_t$, the $t\anti t$ final state has  $\sig(gg\to\aii)\br(\aii\to t \anti t)> 0.01\pb$ for $\maii<550\gev$, implying $>200$ events for $L=20\fbi$. A study is needed to determine if this would be observable in the presence of the $t\anti t$ continuum background.  No doubt, efficient $b$ tagging and reconstruction of the $t\anti t$ invariant mass in, say, the single lepton final state would be needed. For $\maii<2m_t$, the $X=\ai\hii$ final state with both $\ai$ and $\hii$ decaying to $b\anti b$ might be visible above backgrounds.   However, a dedicated study of this particular decay 
mode is still lacking. Similar remarks apply in the case of the $\hiii$ where the possibly visible final states are $t\anti t$ for $\mhiii>2m_t$ and $\hi\hii$ for $\mhiii<2m_t$.  For both the $\aii$ and $\hiii$, $\sig\br(X)$ is substantial for $X=\cnone\cnone$, but to isolate this invisible final state would require an additional photon or jet tag which would reduce the cross section from the level shown.  

\begin{figure}[t]
\hspace*{-4mm}
\includegraphics[width=0.5\textwidth]{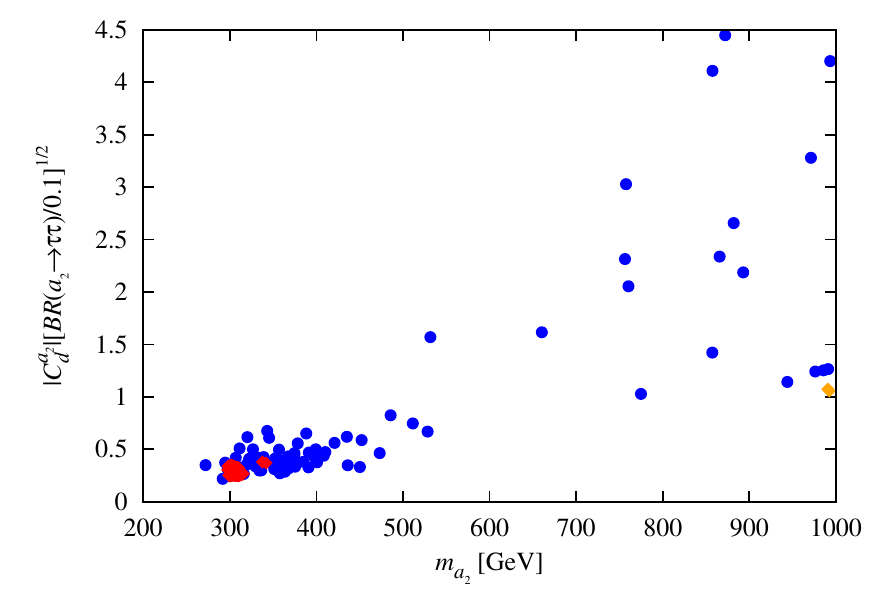}
\hspace*{-6mm}
\includegraphics[width=0.5\textwidth]{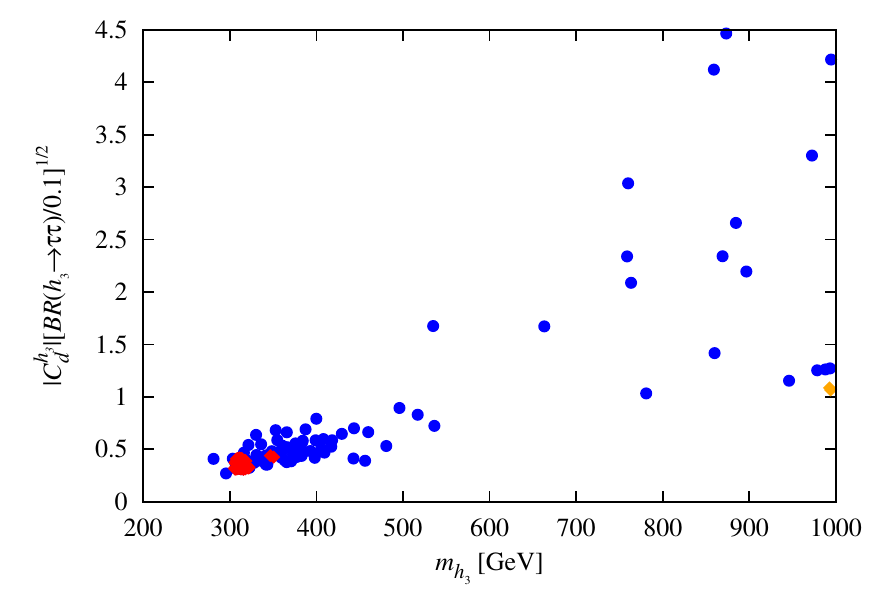}
\caption{$C_d^{\aii,\hiii}({\rm eff})$, see Eq.~(\ref{cdeffdef}), vs. $\maii$ and $\mhiii$ for $gg\to \aii,\hiii\to\tauptaum$.}
\label{cdeffvsm}
\end{figure}

A final possible detection mode is $gg\to\aii,\hiii\to\tauptaum$.  For this case we plot in Fig.~\ref{cdeffvsm} the effective down-quark coupling, $C_d^{\aii,\hiii}(\rm eff)$ vs. $\maii$ and $\mhiii$, where we define 
\beq
C_d^{\aii,\hiii}({\rm eff})=|C_d^{\aii,\hiii}|\left[ {\br (\aii,\hiii\to\tauptaum)\over 0.1 }\right ] ^{1/2}
\label{cdeffdef}
\eeq
and where $0.1$ is a reference value of $\br(H,A\to\tauptaum)$ implicit in the MSSM limit plots discussed below.
 Noting that $\maii\simeq \mhiii$ and the fact that the two plots are nearly identical shows that we may sum the $\aii$ and $\hiii$ signals together in the same manner as the $H$ and $A$ signals are summed together in the case of the analogous plot of $\tanb$ vs. $m_A\simeq m_H$ in the case of the MSSM.
Limits from CMS $4.6\fbi$ data \cite{Chatrchyan:2012vp} are of order $C_d^{\aii,\hiii}({\rm eff})\lsim 7-8$ for $\maii\simeq\mhiii\in[150,220]\gev$ rising rapidly to reach $\sim 50$ at degenerate mass of order $500\gev$.  A dedicated study is needed to determine the precise luminosity for which LHC detection or meaningful limits will become possible for $C_d^{\aii,\hiii}({\rm eff})\lsim 1$ (as relevant for $\maii,\mhiii<550\gev$).  Even though Higgs cross sections from $gg$ fusion increase, relative to $\rts=8\tev$, for $\rts=14\tev$ quite high luminosity will be needed.  Currently, for example, the  CMS limit from $10\fbi$ of data at $\maii\simeq \mhiii\sim 300\gev$ is of order 18, and this amplitude level limit will only improve statistically by $1/L^{1/4}$. Even accounting for the $\rts=14\tev$ cross section increase, very significant improvements in the sensitivity of this analysis will be needed.

\begin{figure}[h]\centering
\includegraphics[width=0.7\textwidth]{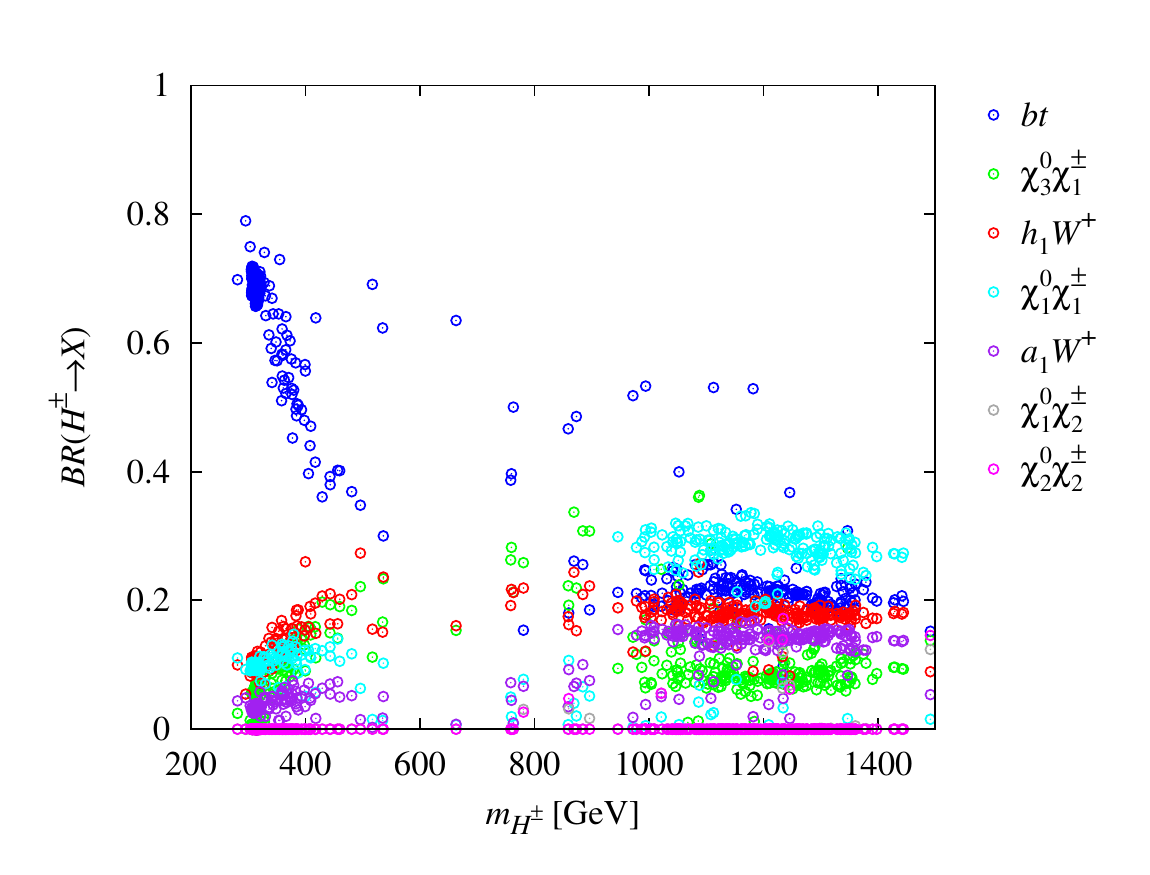}
\caption{Decay branching ratios of the charged Higgs bosons. 
\label{chargedhiggs}}
\end{figure}

The branching ratios for the $\hpm$ are plotted in Fig.~\ref{chargedhiggs}, Prospects for its discovery at masses for which $\hp\hm$ production has substantial cross section appear to be promising in the $bt$ final state  provided reconstruction of the $bt$ mass is possible with good efficiency and one or more $b$ tags are sufficient to reject SM background.  Also very interesting would be detection of  $\hpm\to \hi\wpm$ in the $\hi\to b\anti b$ final state using mass reconstruction for the $b\anti b$ and a leptonic trigger from the $\wpm$ to reject backgrounds.  This channel could prove especially essential in order to detect the $\mhi\sim 98\gev$ Higgs at the LHC and verify the $98+125\gev$ Higgs scenario.

\vspace*{-.2in}
\subsection{\boldmath Higgses from neutralino decays}
\vspace*{-.1in}

\begin{figure}[h]
\hspace*{-7mm}
\includegraphics[width=0.535\textwidth]{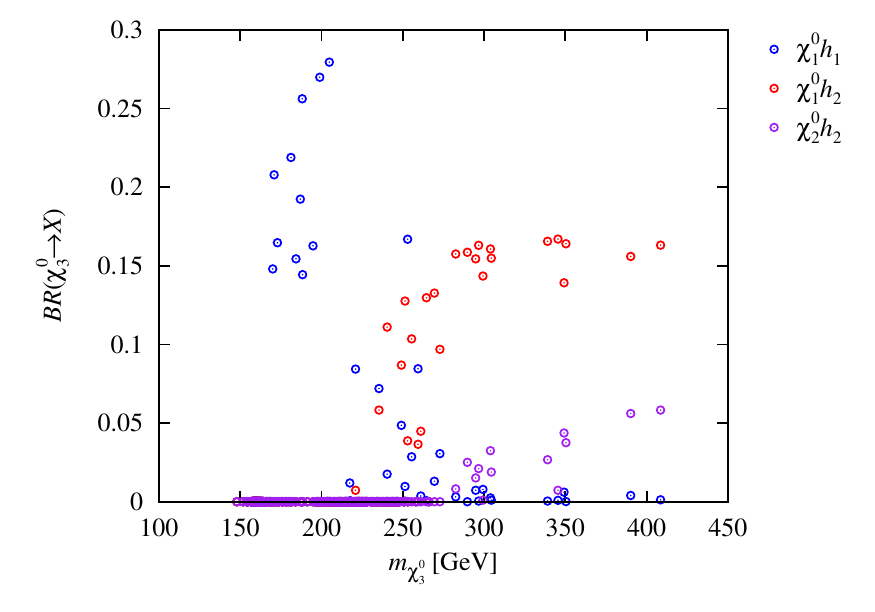}
\hspace{-8mm}
\includegraphics[width=0.535\textwidth]{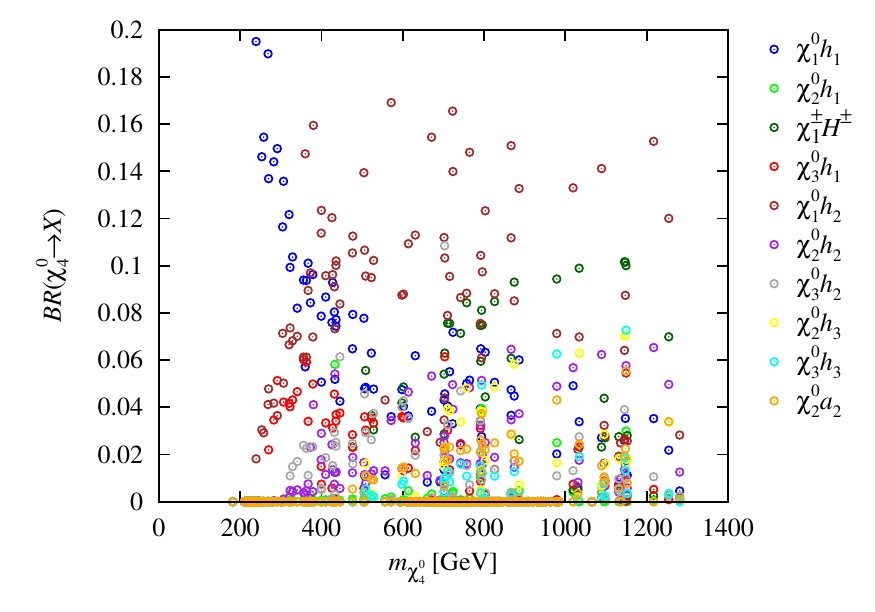}
\hspace*{-7mm}
\includegraphics[width=0.535\textwidth]{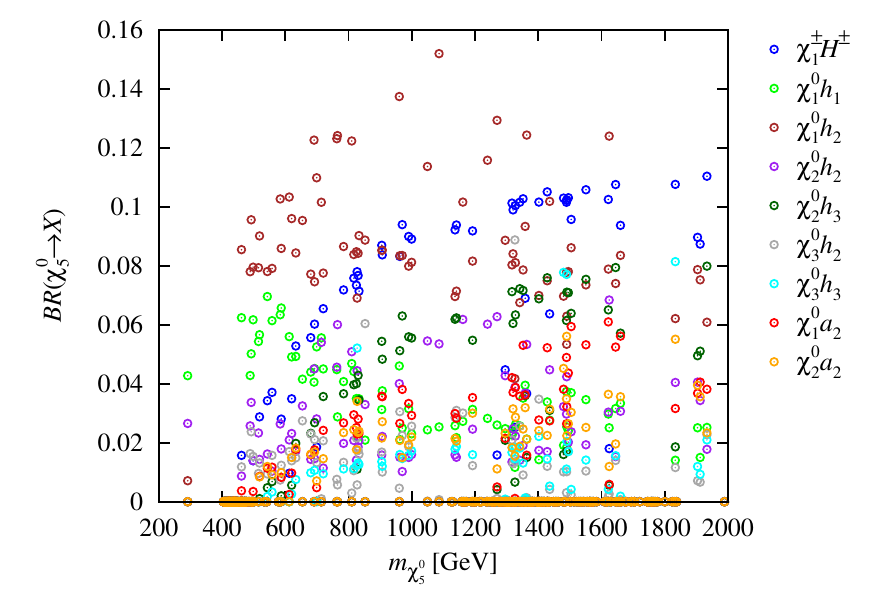}
\hspace{-8mm}
\includegraphics[width=0.535\textwidth]{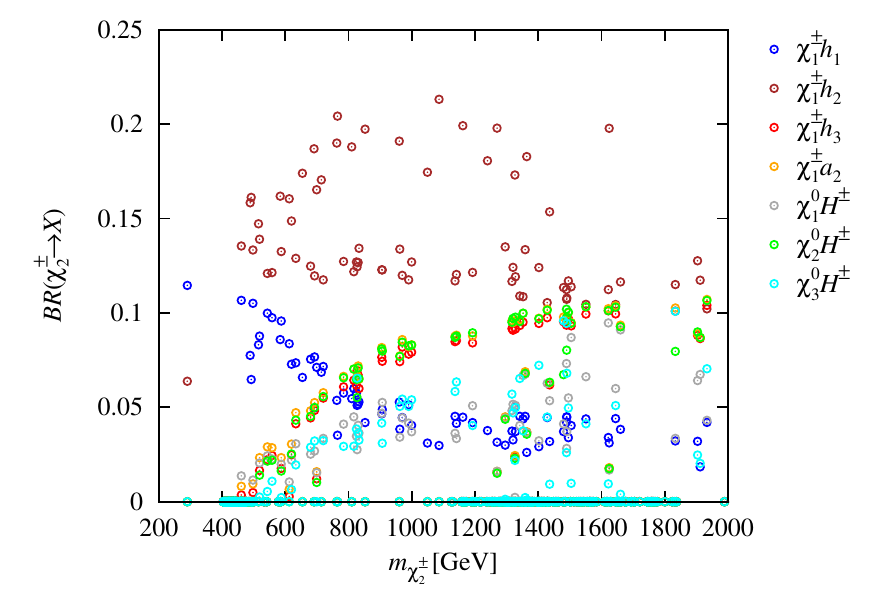}
\vspace*{-4mm}
\caption{Branching ratios for neutralino and chargino decays into final states containing a Higgs boson for the $98+125\gev$ LEP-LHC Higgs scenarios.
\label{plots10}}
\end{figure}

Given that cascades from gluinos/squarks will have low event rate as a result of  the large $\mgl$ and $\msq$ masses predicted  and the rather low $\cpmone$ and $\cnone$ masses typical of the NMSSM scenarios we discuss, prospects for detecting chargino pair production and neutralino+chargino production would appear to be better, although one is faced with cross sections that are electroweak in size. Of particular interest is whether some of the Higgs bosons can be detected via ino-pair production. To assess the possibilities, we present in Fig.~\ref{plots10} the branching ratios for the decay of the neutralinos and charginos to lighter inos plus a Higgs boson.  A brief summary of the results shown is in order.  First, decays to the $\ai$ are not shown since they have very low branching ratios due to the singlet nature of the $\ai$. The only decay with branching ratio to the $\aii$ above 0.1 is $\cpmtwo\to\cpmone \aii$ with $\mcpmtwo\gsim 1.4\tev$ (beyond LHC reach via electroweak production). 
In contrast, prospects for the all important $\hi$ are quite good, with $\br(\cnthree,\cnfour\to\cnone\hi)$ and $\br(\cpmtwo\to \cpmone\hi)$ being quite substantial (\ie\ $>0.1$) at lower values of $\mcnthree,\mcnfour$ and $\mcpmtwo$, respectively.  Decays of $\cnthree,\cnfour,\cnfive$ to $\cnone\hii$ all have $\br>0.1$ once $\mcnthree,\mcnfour,\mcnfive$ are $\gsim 250,400,500\gev$, respectively. Similarly, $\br(\cpmtwo\to \cpmone\hii)>0.1$ for $\mcpmtwo\gsim 500\gev$. Since the charged Higgs has $\mhpm>300\gev$, decays to it, although present for the $\cnfour$, $\cnfive$ and $\cpmtwo$, do not have $\br>0.1$ until $\mcnfour,\mcnfive,\mcpmtwo\gsim 1.1,1.3,1.3 \tev$, respectively.

\vspace*{-.2in}
\subsection{Linear Collider and Photon Collider Tests}
\vspace*{-.1in}

\begin{table}[b]
\caption{Higgs masses and LSP mass in GeV for the three scenarios for which we plot $\epem$ cross sections in Fig.~\ref{LC}.  Also given are $\omghsq$, the singlino and Higgsino percentages and $R_{gg}^{\hii}(\gam\gam)$. Scenarios I) and III) have $\omghsq$ in the WMAP window, with I) being typical of the low-$\mcnone$ scenarios and III) being that with smallest $\mhiii$ in the large-$\mcnone$ group of points in the WMAP window.  Scenario II) is chosen to have $\maii$ and $\mhiii$ intermediate between those for scenario I) and III), a region for which $\omghsq$ is substantially below $0.1$.\label{LCtab}}
\begin{center}
\begin{tabular}{|c|c|c|c|c|c|c|c|c|c|c|c|}
\hline
Scenario & $\mhi$ & $\mhii$ & $\mhiii$ & $\mai$ & $\maii$ & $\mhpm$ & $\mcnone$ & $\omghsq$  & LSP singlino & LSP  Higgsino  & $R_{gg}^{\hii}(\gam\gam)$ 
\\
\hline
I & 99 & 124 & 311 & 140 &  302 & 295 & 76 & 0.099  & 18\% & 75\% & 1.62\\
II  & 97 & 124 & 481 & 217 & 473 & 466 & 92 & 0.026 & 20\% & 74 \% & 1.53\\
III & 99 & 126 & 993 & 147 & 991 & 989 & 115 & 0.099 & 75\% & 25\% & 1.14 \\
\hline
\end{tabular}
\end{center}
\end{table}

\begin{figure}[t]\begin{center}
\hspace*{8mm} scenario I\hspace{63mm} scenario II\\
\includegraphics[width=0.48\textwidth]{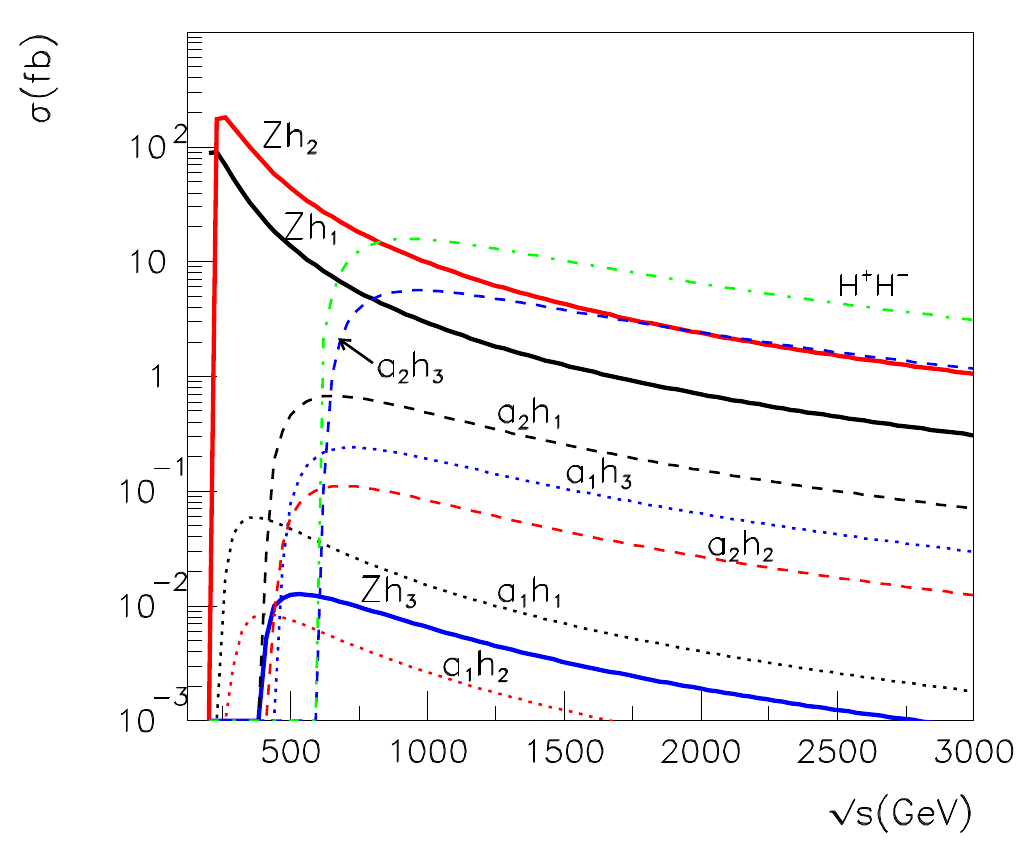}
\includegraphics[width=0.48\textwidth]{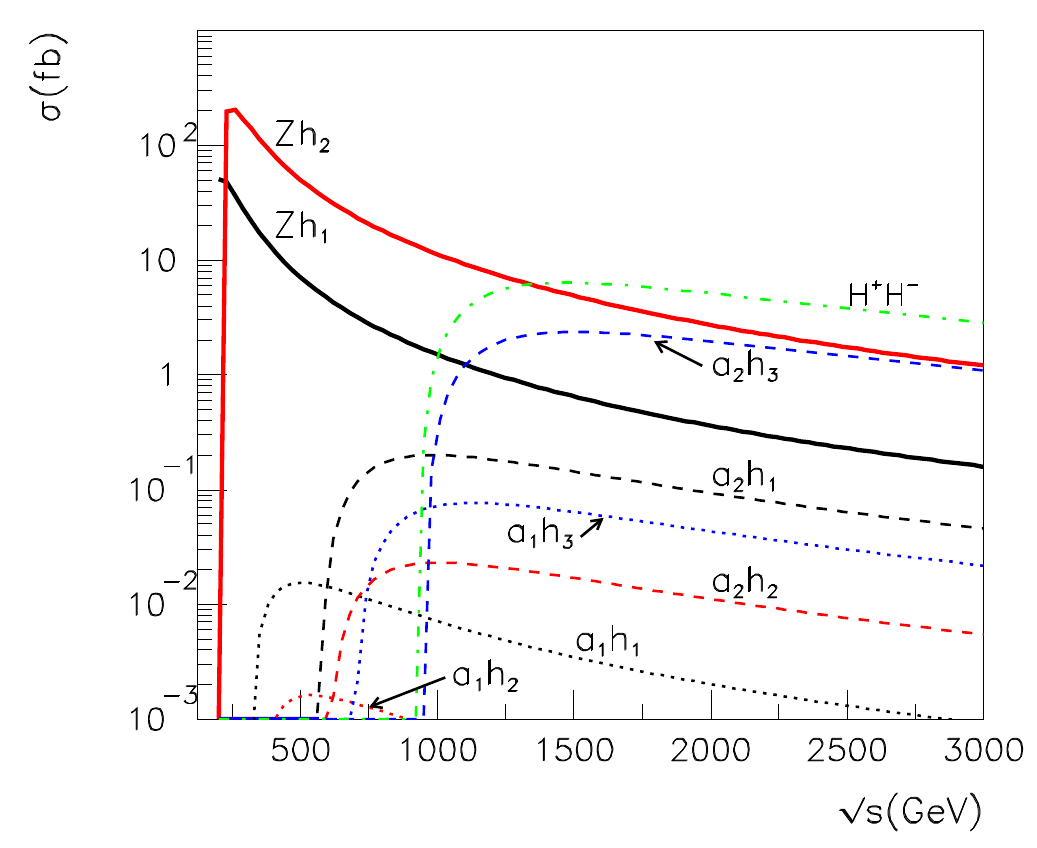}
\hspace*{10mm} scenario III\\
\includegraphics[width=0.5\textwidth]{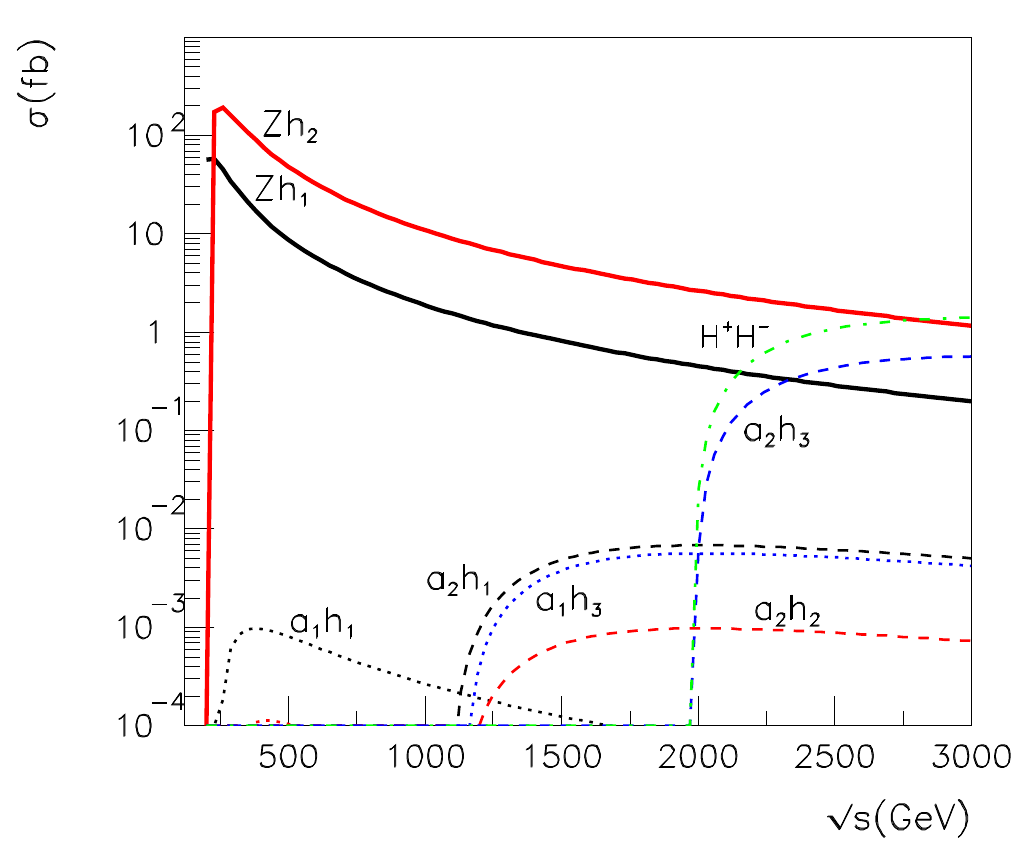}\hspace*{-.13in}
\end{center}
\vspace*{-4mm}
\caption{Cross sections for Higgs production at an $e^+e^-$ collider, as functions of the center-of-mass energy $\sqrt{s}$, for three illustrative mass spectra as tabulated in Table~\ref{LCtab}.
\label{LC}}
\end{figure}

An $e^+e^-$ collider would be the ideal machine to produce the additional Higgs states and resolve the scenario. Production cross sections for the various Higgs final states are shown in Fig.~\ref{LC} for the three illustrative scenarios specified in Table~\ref{LCtab} taken from our NMSSM scans. The first plot is for a WMAP-window scenario with  $\mcnone\sim 76\gev $ and light Higgs bosons. The third plot is for the point  in region B with smallest $\mhiii$, for which $\maii,\mhiii,\mhpm$ are all around $1\tev$. The second plot is for a sample scenario with Higgs masses that are intermediate, as only possible if $\omghsq$ lies below the WMAP window. With an integrated luminosity of $1000\fbi$, substantial event rates for many $Z$+Higgs and Higgs pair final states are predicted.  Of course, $Z\hi$ and $Z\hii$ production have the largest cross sections and lowest thresholds.  The next lowest thresholds are for $\ai\hi$ production, but the cross sections are quite small, $<0.1,0.01,0.001\fb$, respectively.  The $\ai\hii$ cross sections are even smaller. Next in line are $\ai\hiii$, $\aii\hi$ and $\aii\hii$, with $\aii\hi$ having thresholds $>400,600,1190\gev$ for scenarios I), II) and III), respectively, as well as having the largest cross section, peaking at $\sigma>0.7,0.2,0.007\fb$ for the three respective scenarios.   Production of $\aii\hiii$ and $\hp\hm$ have thresholds $>620,950,2000\gev$, respectively, but have much larger cross sections, that for $\hp\hm$ being $>16.6,6.3,1.4\fb$ at the peak, for the three respective scenarios.

In the $\epem$ collider case, it would be easy to isolate signals in many final states.  For example, in the case of Higgs pairs, final states such as $(t\anti t)(t\anti t)$, $(\cnone\cnone)(t\anti t)$ and so forth could be readily identified above background.  Observation of the $(\cnone\cnone)(\cnone\cnone)$ final states would require a photon tag and would thus suffer from a reduced cross section. Associated $Z$+Higgs, with Higgs decaying to $t\anti t$ or $\cnone\cnone$ would be even more readily observed.

Another future collider that would become possible if an $e^+e^-$ (or $e^-e^-$) collider is built is a $\gam\gam$ collider where the $\gam$'s are obtained by backscattering of laser photons off the energetic $e$'s.  For a recent summary see~\cite{Gronberg:2012yj} and references therein. A huge range of energies is possible for such a $\gam\gam$ collider, ranging from low to high center of mass energies depending upon the center of mass energy of the underlying electron collider. A $\gam\gam$ collider based on $e^-e^-$ collisions can even be considered as a stand-alone machine that could be built before an $\epem$ collider, especially if high  $\rts_{\gam\gam}$ is not needed. Typically, the largest $\rts_{\gam\gam}$ that is possible with large instantaneous $\gam\gam$ luminosity is of order $0.8 \rts_{\epem}$. That $\gam\gam\to$Higgs is an effective way to study a SM Higgs boson has been well established~\cite{Asner:2001ia,Velasco:2002vg,Asner:2003hz}.  For low Higgs masses, the required electron collider could have energy of order $m_{\rm Higgs}/0.8$. 

\begin{figure}[t]
\begin{center}
\includegraphics[width=0.5\textwidth]{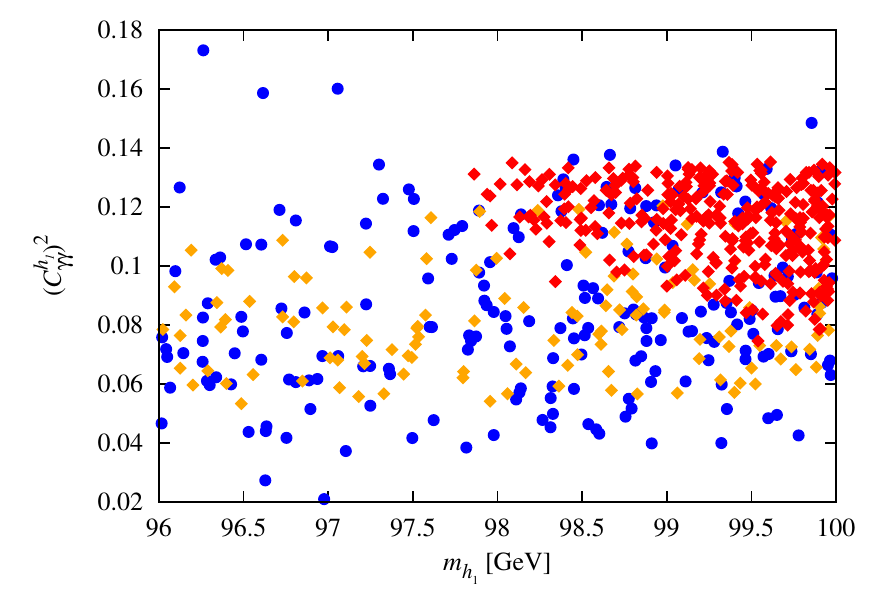}
\hspace{-5mm}
\includegraphics[width=0.5\textwidth]{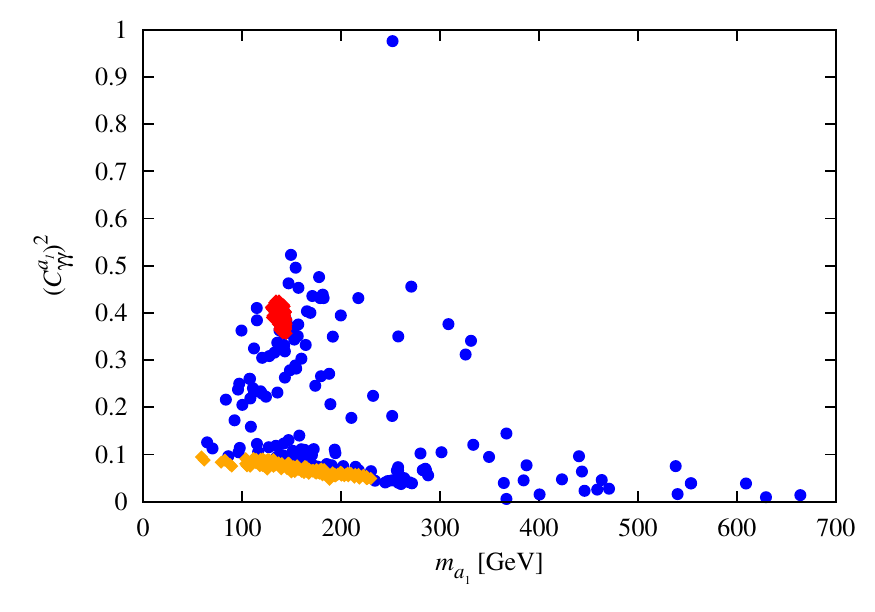}
\includegraphics[width=0.5\textwidth]{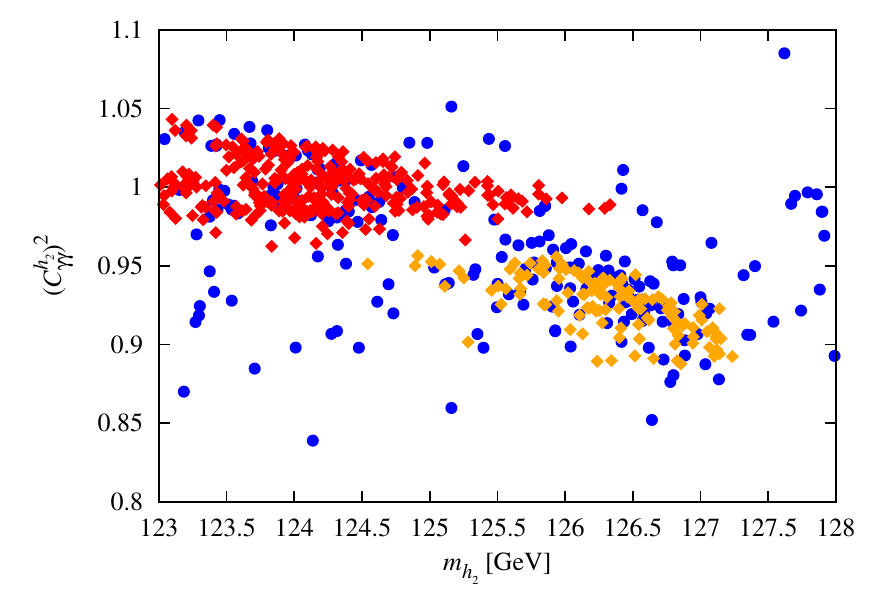}
\hspace{-5mm}
\includegraphics[width=0.5\textwidth]{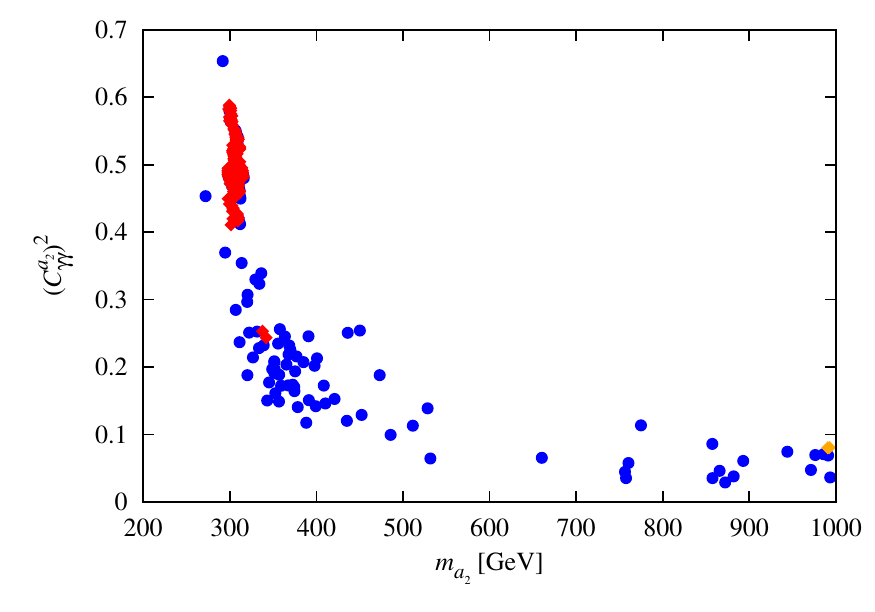}
\includegraphics[width=0.5\textwidth]{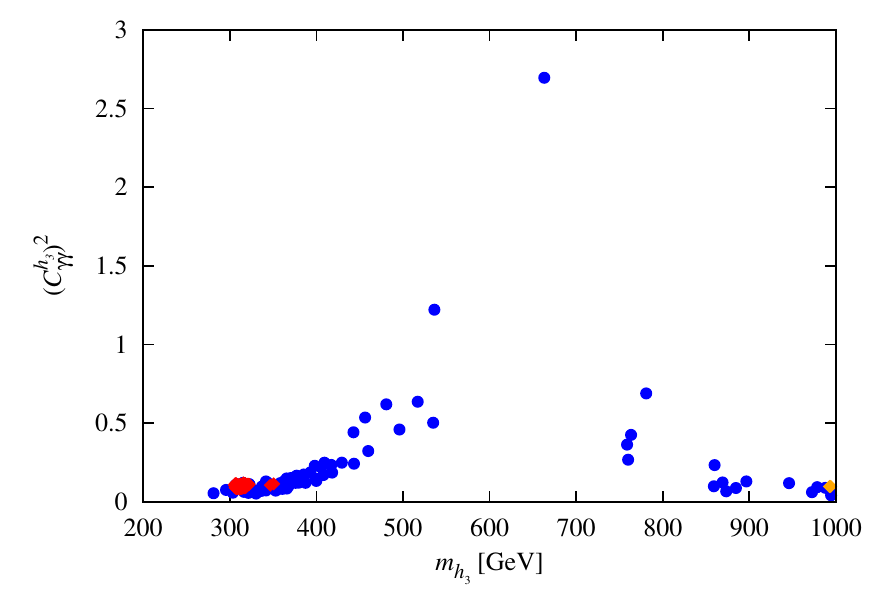}
\caption{$(C^{h}_{\gam\gam})^2$ as a function of $m_h$ for $h=\hi,\hii,\hiii,\ai,\aii$.}
\label{gamgamcoups}
\end{center}
\end{figure}

In the present context, it is of interest to assess the extent to which a $\gam\gam$ collider would be able to study the neutral NMSSM Higgs bosons.  This is determined by the ratio of the $\gam\gam$ coupling squared of the given Higgs boson to that of the SM Higgs. In Fig.~\ref{gamgamcoups} we present plots of $(C^{h}_{\gam\gam})^2$ as a function of $m_h$ for $h=\hi,\hii,\hiii,\ai,\aii$ for masses below $1\tev$. The fairly SM-like $\hii$ at $\sim 125\gev$ can be studied easily at such a collider since its $\gam\gam$ coupling is close to SM strength.  
For example, at an $e^-e^-$ collider with the optimal $E_{ee}=206\gev$, a $125\gev$ SM Higgs has a cross section of $200\fb$.  After two years of operation, equivalent to $L=500\fbi$, one can measure the $b\anti b,\wp\wm,\gam\gam$ partial widths with accuracies of $\Delta\Gamma(b\anti b,\wp\wm,\gam\gam)/\Gamma(b\anti b,\wp\wm,\gam\gam)\sim 0.015,0.04,0.06$, respectively~\cite{Velasco:2002vg} (see also~\cite{Asner:2001ia,Asner:2003hz}). 

Even though the $\hi$ and $\ai$ are largely singlet, both have $\gam\gam$ couplings-squared that are often of order $0.1\times $SM and above (at the same mass). In part, this is because even singlets couple to $\gam\gam$ through a Higgsino-like chargino loop using the singlet-Higgsino-Higgsino coupling that arises from the $\lam \what S \what H_u \what H_d$  term in the superpotential.  Indeed, this coupling becomes stronger as $\lam$ is increased.  Of course, it is important to note that the modest values of $\mueff$ (see Fig.~\ref{plots5}) that characterize many of our scenarios imply that the lightest chargino is largely Higgsino-like and has low mass (see Fig.~\ref{plots5.5}), for which the Higgsino-chargino loop is less suppressed. Even for $\gam\gam$ coupling-squared of order $0.1\times$SM, with sufficient integrated luminosity observation of the $\hi$ and $\ai$ would be possible.  For example, for suitably chosen $E_{ee}$, the above SM Higgs rates multiplied by $0.1$ would roughly apply for $\mhi\sim 98\gev$ or  $\mai < 300\gev$, from which it is clear that the $b\anti b$ final state would be easily observable with $L=500\fbi$ and one could measure the partial width with an accuracy of order $5\%$. Even the $\hiii$ and $\aii$ would be observable for $\maii< 500\gev$, again assuming appropriately optimal $E_{ee}$ for the given $\mhiii$ or $\maii$ and $L=500\fbi$.

This raises the question of whether or not a $\gam\gam$ collider with adjustable (as is straightforward) $\rts_{\gam\gam}$ in the $98\gev$ range would be a good next step for high energy physics.  It would have the advantage of allowing important detailed studies of the $\hii$ (or any SM-like Higgs boson with mass of $125\gev$) while testing for the presence of the $\hi$.  With adjustable $\rts_{\gam\gam}$ and $L \geq 500\fbi$, the $\hiii,\ai,\aii$, or any other light Higgs boson with significant (even if somewhat suppressed) $\gam\gam$ coupling, would be observable as well.

\vspace*{-.2in}
\subsection{\boldmath A $\mupmum$ Collider}
\vspace*{-.1in}

\begin{figure}[b]
\begin{center}
\includegraphics[width=0.5\textwidth]{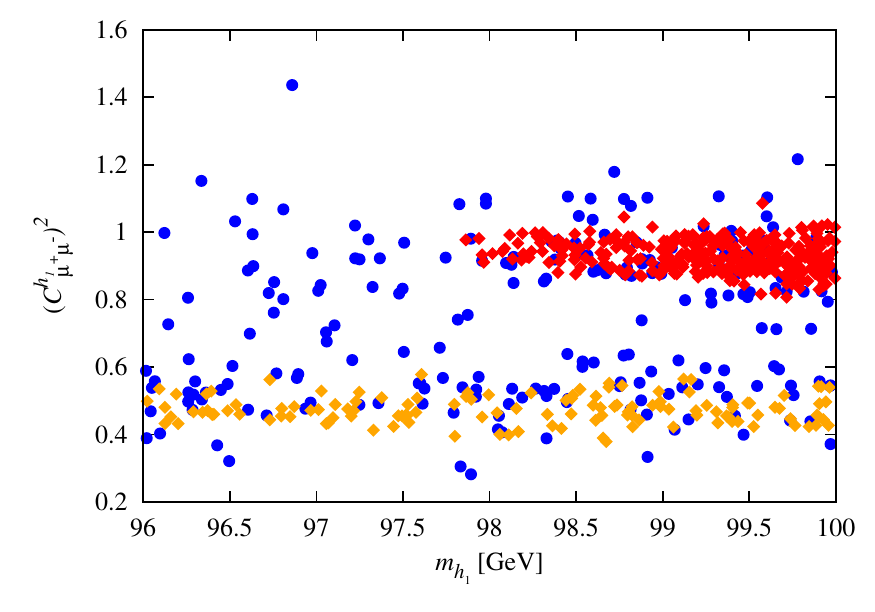}
\hspace{-5mm}
\includegraphics[width=0.5\textwidth]{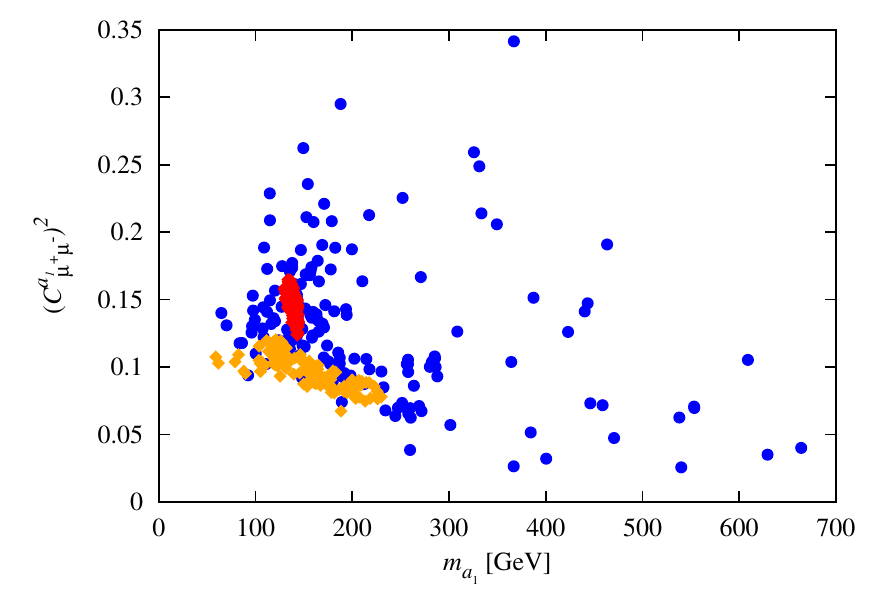}
\includegraphics[width=0.5\textwidth]{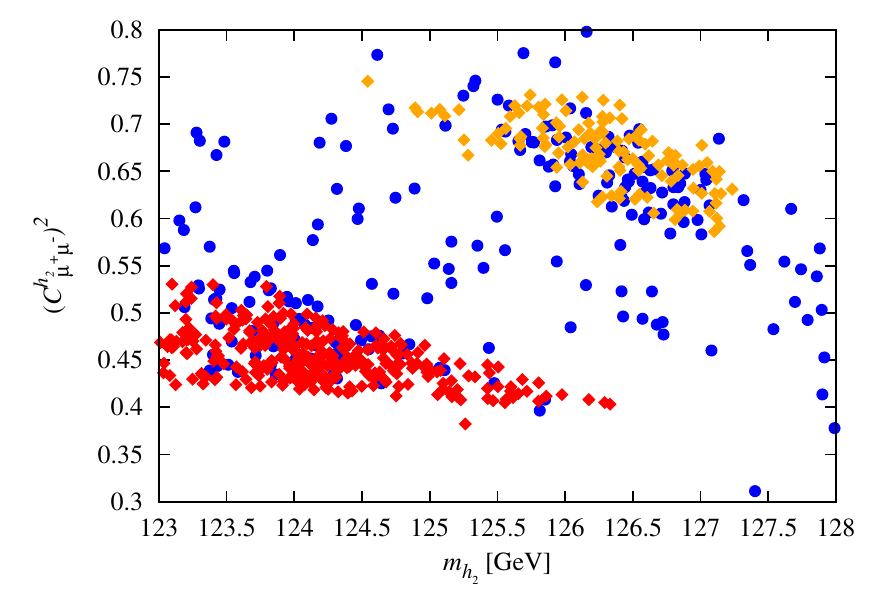}
\hspace{-5mm}
\includegraphics[width=0.5\textwidth]{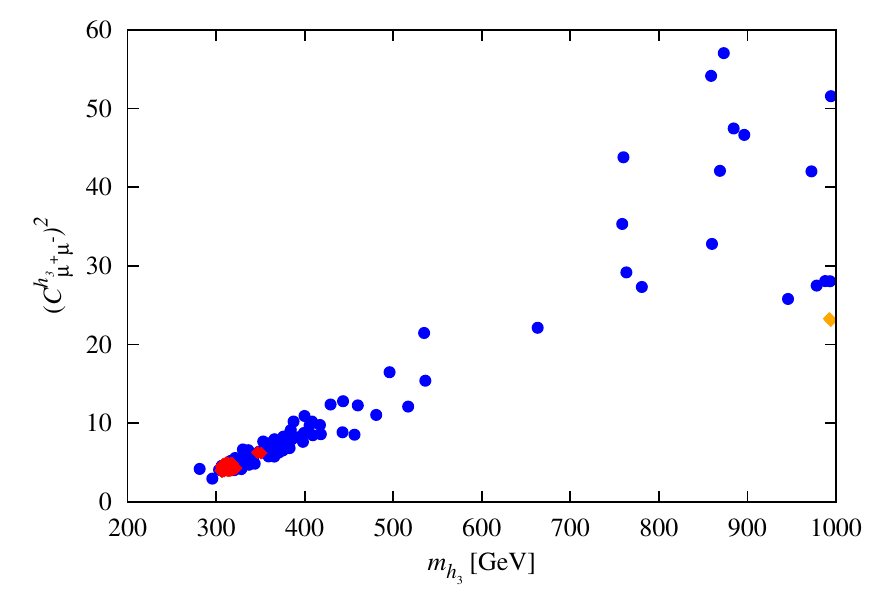}
\end{center}
\caption{Reduced $\mupmum$ couplings squared for $\hi,\hii,\hiii,\ai$.\protect \label{mucoup}}
\end{figure}

A muon-collider with $\rts$  close to the Higgs mass in question would be a particularly ideal machine to study any Higgs boson with $\mupmum$ coupling that is not too different from that of a SM Higgs boson of similar mass. Thus, in Fig.~\ref{mucoup} we present plots of $(C_{\mupmum}^h)^2$ as a function of $m_h$ for $h=\hi,\hii,\hiii,\ai$, that for the $\aii$ being essentially identical to the $h=\hiii$ case.  We see that prospects are really quite good for the $\hi$ as well as the $\hii$.  In addition, the WMAP-window $\ai$ points, all of which lie at relatively low mass, can be probed as well.  As for the $\hiii$ (and the $\aii$), the low-$\mcnone$ region points with low $\mhiii$ ($\simeq \maii$) have nicely enhanced $(C^{\hiii}_{\mupmum})^2$ ( $\simeq (C^{\aii}_{\mupmum})^2$).  A muon collider would be ideal for probing such scenarios.  Additional experimental evidence for this $98+125\gev$ Higgs scenario from other machines would provide strong motivation for the muon collider.

\section{Conclusions}

To summarize, we have emphasized the possibility that both the LEP excess in the $b\anti b$ final state at $M_{b\anti b}\sim 98\gev$ and the LHC Higgs-like signal at $\sim 125\gev$ with an enhanced rate in the two-photon final state can be explained in the context of the NMSSM.  
The NMSSM scenarios of this type have many attractive features.  We have particularly emphasized the fact that the $\hi$ could eventually be observed at the LHC in $gg,{\rm VBF} \to \hi\to b\anti b$.  We urge the ATLAS and CMS collaborations to give attention to this possibility. 

The $98+125\gev$ Higgs scenarios have important implications for the other Higgs bosons and for supersymmetric particles. 
If we focus only on the subset of these  scenarios that have relic density in the WMAP window, then there are two separate regions of NMSSM parameter space that emerge.  One  region (A) is characterized by small $\mcnone\sim 75\gev$ and low masses for many of the Higgs bosons and superpartners, including $\mstopi$ as low as $197\gev$. The second  region (B) is characterized by larger $\mcnone\in [93,150]\gev$ and much larger mass scales for the heavier Higgs bosons and supersymmetric particles.  For this latter region, one finds $\mai\in[100,200]\gev$, $\mcpmone\in[170,230]\gev$, $\maii\simeq\mhiii\simeq \mhpm\in[1,1.4]\tev$, $\mstopi\in[1.9,2.8]\tev$,  $\msq,\mgl\in[3,5]\tev$ and $\tanb\in[5,7]$.
Clearly this latter region leaves little hope for LHC detection of the colored particles and experimental probes would need to focus on the gauginos and lighter Higgs bosons.  It is further associated with rather modest values for the enhancement of the $125\gev$ Higgs signal in the $\gam\gam$ channel.
Information related to the prospects for Higgs and superparticle detection for the two regions (A) and (B) at an $\epem$, $\gam\gam$ or  $\mupmum$ collider are summarized.

\section*{Acknowledgements} 

The work of JFG and YJ was supported by US DOE grant DE-FG03-91ER40674, that of JHS the U.S. DOE grant No. DE-FG03-92-ER40701, and that of SK and GB by
IN2P3 under contract PICS FR--USA No.~5872. UE acknowledges partial support from the French ANR~LFV-CPV-LHC, ANR~STR-COSMO and the European Union FP7 ITN INVISIBLES (Marie Curie
Actions,~PITN-GA-2011-289442).
GB, UE, JFG, SK, and JHS acknowledge the hospitality and the inspiring working atmosphere  
of the Aspen Center for Physics which is supported by the National Science Foundation Grant No.\ PHY-1066293.


\end{document}